\documentclass[printer]{aa}
\usepackage{times}
\usepackage{natbib}
\usepackage{graphicx,longtable,cancel}
\usepackage{caption}
\usepackage{subfig}
\usepackage{booktabs}
\usepackage{tablefootnote}
\usepackage{amsmath,geometry}
\usepackage{float,lscape}
\usepackage{xspace}
\usepackage{verbatim}
\usepackage{txfonts}

\usepackage{supertabular}

\bibpunct{(}{)}{;}{a}{}{,}

\def\pdot {\dot P}
\def\edot {\dot E}

\def\x{\times}
\def\nh{$N_{\rm H}$\xspace}

\def\flux{erg~s$^{-1}$~cm$^{-2}$\xspace}
\def\lum{erg~s$^{-1}$\xspace}

\newcommand{\band}[2]{#1--#2 keV\xspace}
\newcommand{\coord}[8]{R.A.\,=\,${#1}^{\rm h} {#2}^{\rm m} {#3}^{\rm s}.{#4}$, decl.\,=\,${#5}^\circ {#6}' {#7}''.{#8}$} 
\def\psra{PSR\,B0628$-$28\xspace}
\def\psrb{PSR\,B0919$+$06\xspace}
\def\psrc{PSR\,B0114$+$58\xspace}
\def\psrd{PSR\,B1133$+$16\xspace}
\def\psrx{PSR\,B0943$+$10\xspace}
\def\xmm{{\em XMM--Newton}\xspace}

\def\chandra{{\em Chandra}\xspace}

\begin{document}

\title
{A new X-ray look into four old pulsars}

\author{Michela Rigoselli\inst{1,2}, Sandro Mereghetti\inst{1}}

\institute {INAF, Istituto di Astrofisica Spaziale e Fisica Cosmica Milano, via E.\ Bassini 15, I-20133 Milano, Italy
\and
Dipartimento di Fisica G. Occhialini, Universit\`a degli Studi di Milano Bicocca, Piazza della Scienza 3, I-20126 Milano, Italy }

\offprints{m.rigoselli@campus.unimib.it}

\date{Received / Accepted}

\authorrunning{M. Rigoselli \& S. Mereghetti}

\titlerunning{ }

\abstract{We report on the X-ray properties of four rotation-powered pulsars with characteristic ages in the range $0.3-5$ Myr, derived from the analysis of \xmm archival observations.
We found convincing evidence of thermal emission only in the phase-averaged spectrum of  \psrc, that is well fitted by a blackbody with temperature $kT=0.17\pm 0.02$~keV and emitting radius $R=405_{-90}^{+110}$~m, consistent with the size of its polar cap.  
The other three considered pulsars, \psra, \psrb and \psrd, have phase-averaged spectra well described by single power-laws with photon index $\Gamma\sim3$. The $3\sigma$ upper limits on the bolometric luminosity of a possible thermal component with temperature in the range $\sim \! 0.05-2$ keV are $L_{\rm bol} \lesssim 3.2 \times 10^{28}$ \lum and $L_{\rm bol} \lesssim 2.4 \x 10^{29}$ \lum, for \psra and \psrb,  respectively.  On the other hand, we found possible evidence that the pulsed emission of \psra is thermal. 
Two absorption lines at $\sim\!0.22$ keV and $\sim\!0.44$ keV are detected in the spectrum of \psrd. They are best interpreted as proton cyclotron features, implying the presence of multipolar components with a field of a few $10^{13}$ G at the neutron star  polar caps. 
We discuss our results in the context of high-energy emission models of old rotation-powered pulsars.

\keywords{pulsar: general -- pulsars: individual (\psra, \psrb, \psrc, \psrd) -- stars: neutron -- X-rays: stars}}

\maketitle

\section{Introduction}

\setlength{\tabcolsep}{0.5em}
\begin{table*}
\centering \caption{Pulsar properties}
\label{tab:infopulsar}

\begin{tabular}{lcccccccc}
\toprule
Pulsar Name	& $P$	& $\pdot$				& $\edot$			& $\tau$& $B_d$			& DM			& \nh$\,^{\rm a}$		& $d$				\\[5pt]
		& s			& $10^{-15}$~s~s$^{-1}$	& erg~s$^{-1}$		& Myr	& $10^{12}$~G	& cm$^{-3}$~pc	& $10^{20}$~cm$^{-2}$	& kpc				\\[5pt]				
\midrule\\[-5pt]
\psra 	& $1.244$	& $7.12$				& $1.5 \x 10^{32}$	& $2.77$& $3.0$			& $34.42$		& $10$	& $0.32_{-0.04}^{+0.05}\,^{\rm b}$	\\[3pt]
\psrb	& $0.431$	& $13.73$				& $6.8 \x 10^{33}$	& $0.50$& $2.5$			& $27.30$		& $8$	& $1.1_{-0.1}^{+0.2}\,^{\rm b}$		\\[3pt]
\psrc	& $0.101$	& $5.85$				& $2.2 \x 10^{35}$	& $0.28$& $0.8$			& $49.42$		& $15$	& $1.77 \pm 0.53\,^{\rm c}$			\\[3pt]
\psrd	& $1.188$	& $3.73$				& $8.8 \x 10^{31}$	& $5.04$& $2.1$			& $4.84$		& $1.5$	& $0.35 \pm 0.02\,^{\rm b}$			\\[5pt]
\bottomrule\\[-5pt]
\end{tabular}

\raggedright
\scriptsize
$^{\rm a}$ Inferred from the dispersion measure assuming a $10\%$ ionization of the interstellar medium \citep{he13}.\\
$^{\rm b}$ Parallax measurements \citep{ver12}.\\
$^{\rm c}$ Inferred from the dispersion measure \citep{yao17}. Error assumed $30\%$.\\
\end{table*}

X-rays have been detected from more than one hundred  rotation-powered pulsars.
The properties of the X-ray emission,  and in particular the relative strength of the non-thermal and thermal spectral components, depend on the  pulsar age \citep{bec09,har13}. 

Young pulsars have the highest surface temperatures, but their thermal flux is unobservable, being outshined by a much stronger non-thermal emission. 
The latter fades as the pulsars age and slow-down, thus unmasking the underlying thermal emission of the neutron star (NS) surface.

Indeed middle aged pulsars show a mix of non-thermal and thermal X-ray emission. The latter is often modelled with two blackbody components: the cooler have emitting areas consistent with emission from the whole NS surface and temperatures in agreement with the prediction of standard cooling theories. The hotter components have blackbody temperatures typically in the range \band{0.1}{0.3} and emitting radii between a few tens and a few hundreds of meters, implying that the X-rays come from a small region of the NS (i.e. PSR$\,$B0656$+$14, PSR$\,$J0633$+$1746, PSR$\,$B1055$-$52 \citep{del05} and PSR$\,$J1740$+$1000 \citep{kar12a}).
 
X-rays have been detected also in older pulsars, that have cooled down to temperatures too low to significantly emit in the X-ray band. In these objects both non-thermal X-rays of magnetospheric origin and/or emission from hot polar caps have been observed with high confidence only in a few objects (see for example PSR$\,$B1929$+$10 \citep{mis08}). In fact, most old pulsars are rather faint X-ray sources and their study can benefit of the use of maximum likelihood (ML) techniques, as demonstrated for \psrx \citep{her13,mer16} and   PSR$\,$B1822$-$09 \citep{her17}. 

Here, we report a reanalysis of \xmm\ archival data of four  pulsars  with characteristic ages in the range $0.3-5$~Myr based on  ML methods.
The main properties of our targets are summarized in Table~\ref{tab:infopulsar}.

\section{Observations and data reduction}
\label{sec:da}

\begin{table*}
\centering \caption{Journal of \xmm observations}
\label{tab:obslog}
\begin{tabular}{llccccc}
\toprule
Pulsar Name	& Obs. ID  		&  Start time  		   & End time				& \multicolumn{3}{c}{Effective Exposure (ks)}\\[3pt]
 		 	&				&           UT		   &   UT					& \multicolumn{3}{c}{Operative Modes$\,^{\rm a}$ }\\[3pt]				
 		 	&				& 					   &						& pn		& MOS1		& MOS2 					\\[5pt]				
\midrule\\[-5pt]
\psra		& 0206630101	& 2004 Feb 28 02:19:26 & 2004 Feb 28 15:51:17	& 42.15		& 44.67		& 44.69					\\[3pt]
			&				&					   &						& PLW		& FW		& FW					\\[5pt]
\midrule
\psrb		& 0502920101	& 2007 Nov 09 22:16:34 & 2007 Nov 10 09:45:10	& 24.07		& 25.61		& 25.63					\\[3pt]
			&				&					   &						& PLW		& FW		& FW					\\[5pt]
\midrule
\psrc		& 0112200201	& 2002 Jul 09 19:53:07 & 2002 Jul 09 22:23:28	& 5.40		& 5.95		& 5.97					\\[3pt]
			&				&					   &						& FW		& FW		& FW					\\[5pt]
\midrule
\psrd		& 0741140201	& 2014 May 25 12:18:42 & 2014 May 25 19:20:22 	& 17.69		& 19.22		& 19.22					\\[3pt]
			& 0741140301	& 2014 May 31 11:34:51 & 2014 May 31 17:58:11 	& 17.91		& 19.42		& 19.42					\\[3pt]
			& 0741140401	& 2014 Jun 14 07:47:26 & 2014 Jun 14 18:20:46 	& 30.95		& 33.68		& 33.69					\\[3pt]
			& 0741140501	& 2014 Jun 22 07:22:13 & 2014 Jun 22 17:02:13 	& 28.53		& 30.92		& 30.92					\\[3pt]
			& 0741140601	& 2014 Jun 28 10:58:52 & 2014 Jun 28 17:55:32 	& 19.74		& 21.41		& 21.41					\\[3pt]
			&				&					   &						& FW		& SW		& SW					\\[5pt]
\bottomrule\\[-5pt]
\end{tabular}

\raggedright
\scriptsize
$^{\rm a}$ PLW = Prime Large Window  (43 ms); FW = Full Window (pn 73 ms, MOS 2.6 s); SW = Small Window (0.3 s).
\end{table*}

We used the data obtained with the pn and the two MOS cameras, which constitute the EPIC instrument and cover the energy range \band{0.2}{12} \citep{str01,tur01}. 
The medium optical filter was used for \psrc and for the MOS cameras data of \psra . All the other data were obtained using the thin filter. The time resolution of the different operating modes used during  our observations are indicated in Table~\ref{tab:obslog}. Note that while the pn time resolution was adequate to reveal the  pulsations in  all our targets, this is not true for the MOS data, except in the case of \psrd.
The data reduction and analysis were done using version 15 of the Science Analysis System (SAS)\footnote{https://www.cosmos.esa.int/web/xmm-newton/sas} and XSPEC (ver. 12.8.2) for the spectral fits.

We reprocessed the pn data using the SAS task {\it epreject} to reduce the detector noise at the lowest energies. We then removed the  time intervals  of high background by rejecting all the periods with a pn count rate  in the range \band{10}{12} larger than 1.0 cts~s$^{-1}$. The resulting net exposure times for each pulsar are indicated in 
Table~\ref{tab:obslog}.

We used single- and multiple-pixel events for  both  the pn and MOS.
The events detected in the two  MOS were combined into a single data set, and in their  analysis we used averaged exposure map and response files. 

We applied a maximum likelihood (ML) analysis to extract the net (i.e. background-subtracted) source counts used for the spectral and timing analysis. The resulting counts spectra were put in a format compatible with analysis in XSPEC. By exploiting the knowledge of the instrumental point spread function (PSF), this method  provides spectra and light curves with a  number of source counts larger than what is typically  obtained with the traditional method based on ``source'' and ``background'' extraction regions.  Another advantage is that the background is directly measured at the position of the source, contrary to what is done with the traditional analysis in which it is estimated from  ``source free'' regions of the image.

The ML method measures the source counts taking into account \textit{all} the events of the region of interest. 
Specifically,  it finds the source flux and background level which maximise  the probability  of obtaining the number of total counts, $N_{ij}$,  measured in each spatial pixel. 
The values of  $N_{ij}$ follow a  Poissionian probability distribution, with  expectation value $\mu_{ij} = \beta + \sigma \times \text{PSF}_{ij}$, where $\beta$ is the background level, that we assumed to be uniform, $\sigma$ is the number of source counts,  and $\text{PSF}_{ij}$ gives the relative contribution of the source flux in the  pixel $(i,j)$ taking into account the instrumental PSF, which depends on the position in the field of view and on the energy of the event. 
We used  spatial pixels of 1$\times$1 arcsec$^2$ and we analysed separately the data of the pn and of the sum of the two MOS, adopting the respective PSFs, derived from in-flight calibrations\footnote{http://www.cosmos.esa.int/web/xmm-newton/calibration/documentation}, with parameters appropriate for the average energy value in the considered bin.

Following \citet{her17}, we generalized the above method to take into account also the pulse phase information of the events.
In this ``3D-ML'' approach, the events are binned in spatial and phase coordinates and the expectation value of bin $(i,j,k)$ is $\mu_{ijk} = \beta + \sigma_u \times \text{PSF}_{ij} + \sigma_p \times \text{PSF}_{ij} \times \Phi_k$.

Now $\sigma_u$ and $\sigma_p$ represent the source counts for the unpulsed and pulsed components, while $\Phi_k$ is the normalized pulse profile at phase bin $k$. With this approach, and assuming a sinusoidal pulse  (consistent with the data of our targets),  we could simultaneously derive the fluxes (and spectra) for  the unpulsed and the pulsed components. 
We define the pulsed fraction as the ratio between the pulsed and the total counts, $\sigma_p/(\sigma_p+\sigma_u)$.

\section{Results}
\label{sec:res}

\begin{figure*}
	\centering
		\includegraphics[width=7cm]{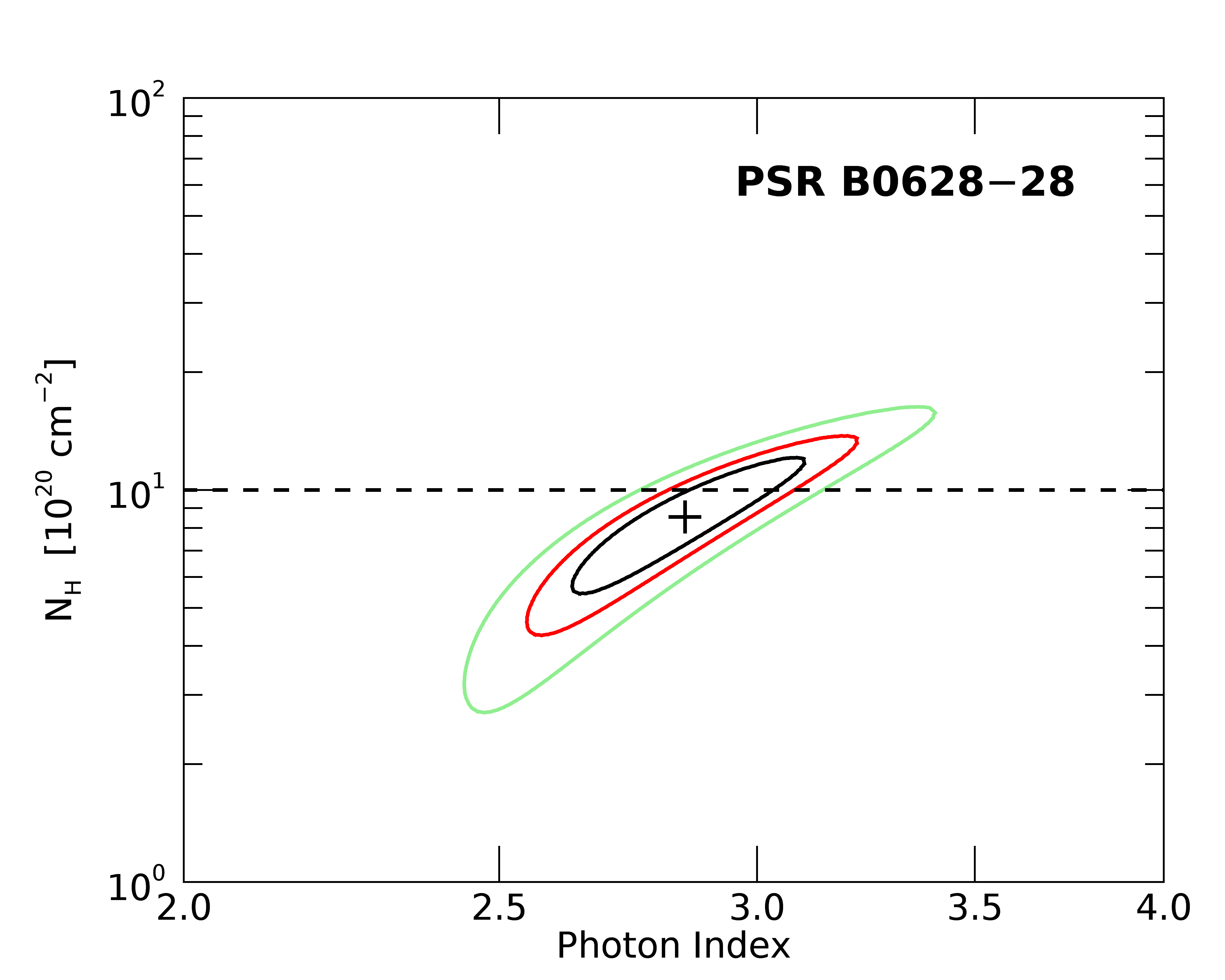}
		\includegraphics[width=7cm]{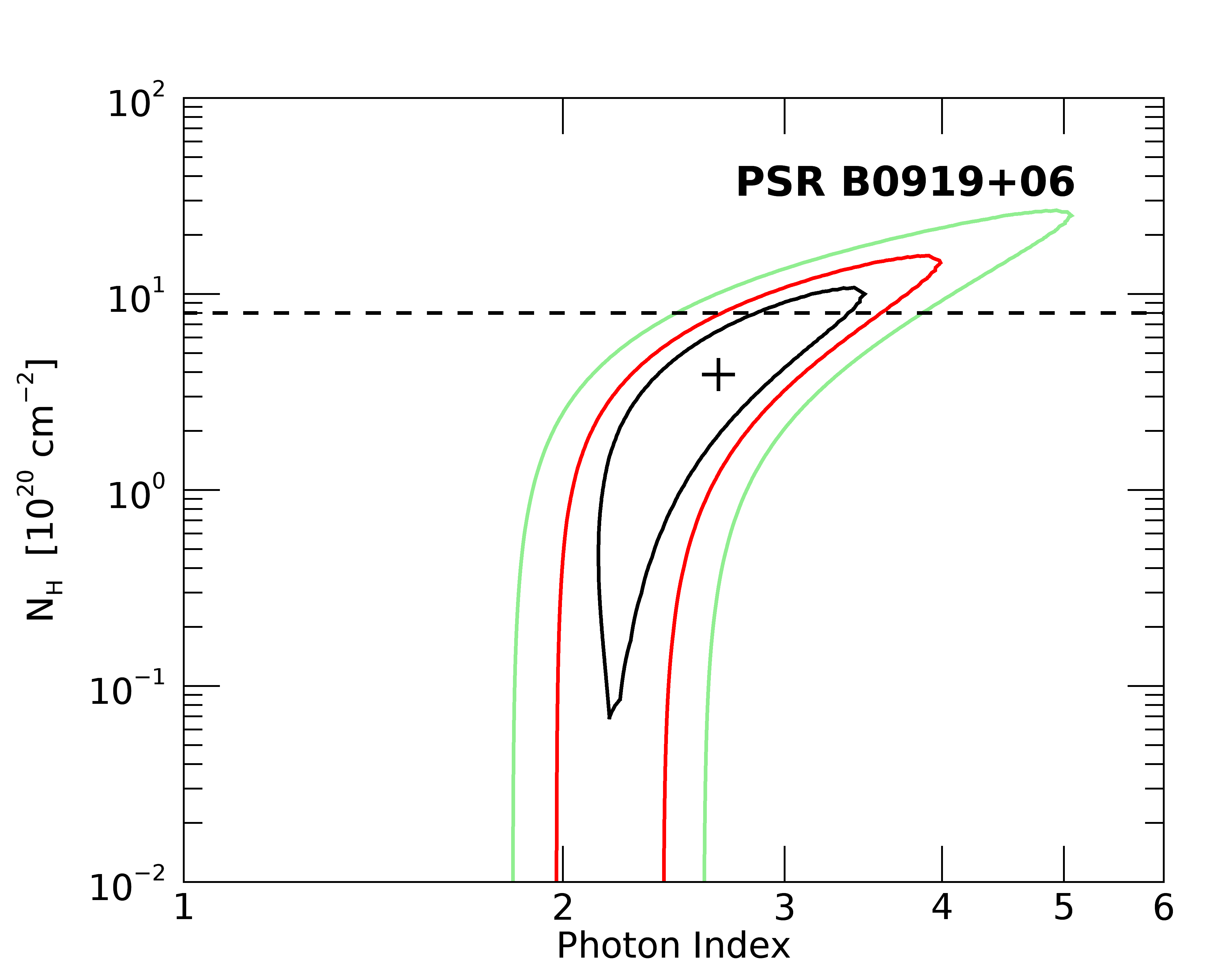}
		\includegraphics[width=7cm]{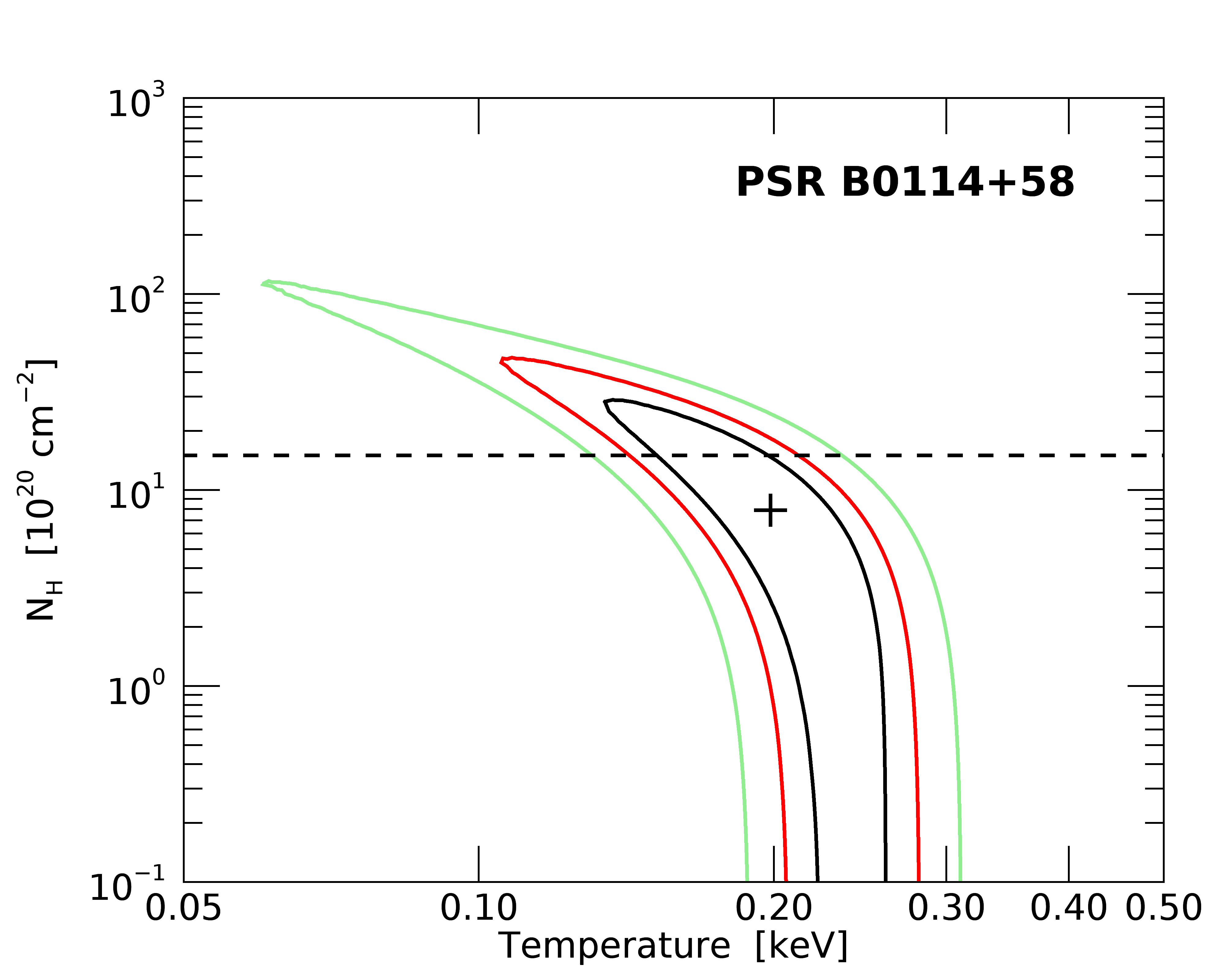}
		\includegraphics[width=7cm]{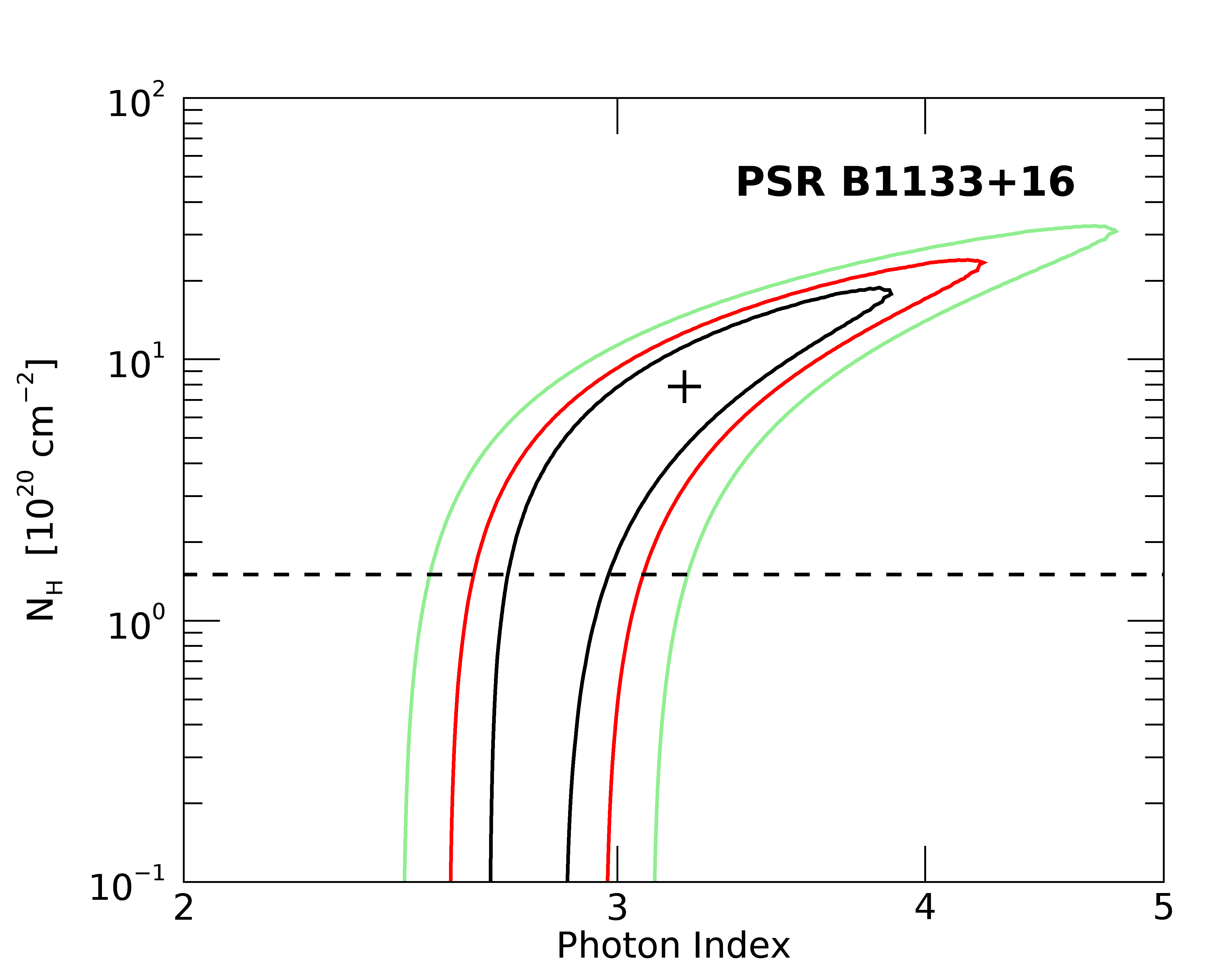}						
	\caption{Confidence regions (1, 2 and 3 $\sigma$) of the best fit parameters for the spectra of (a) \psra, (b) \psrb, (c) \psrc, and (d) \psrd. 	
The best fit values are marked by a plus and the DM-based value of \nh by a dashed line.
\label{fig:nh}}
\end{figure*}

\begin{figure}
	\centering
	\includegraphics[height=8cm]{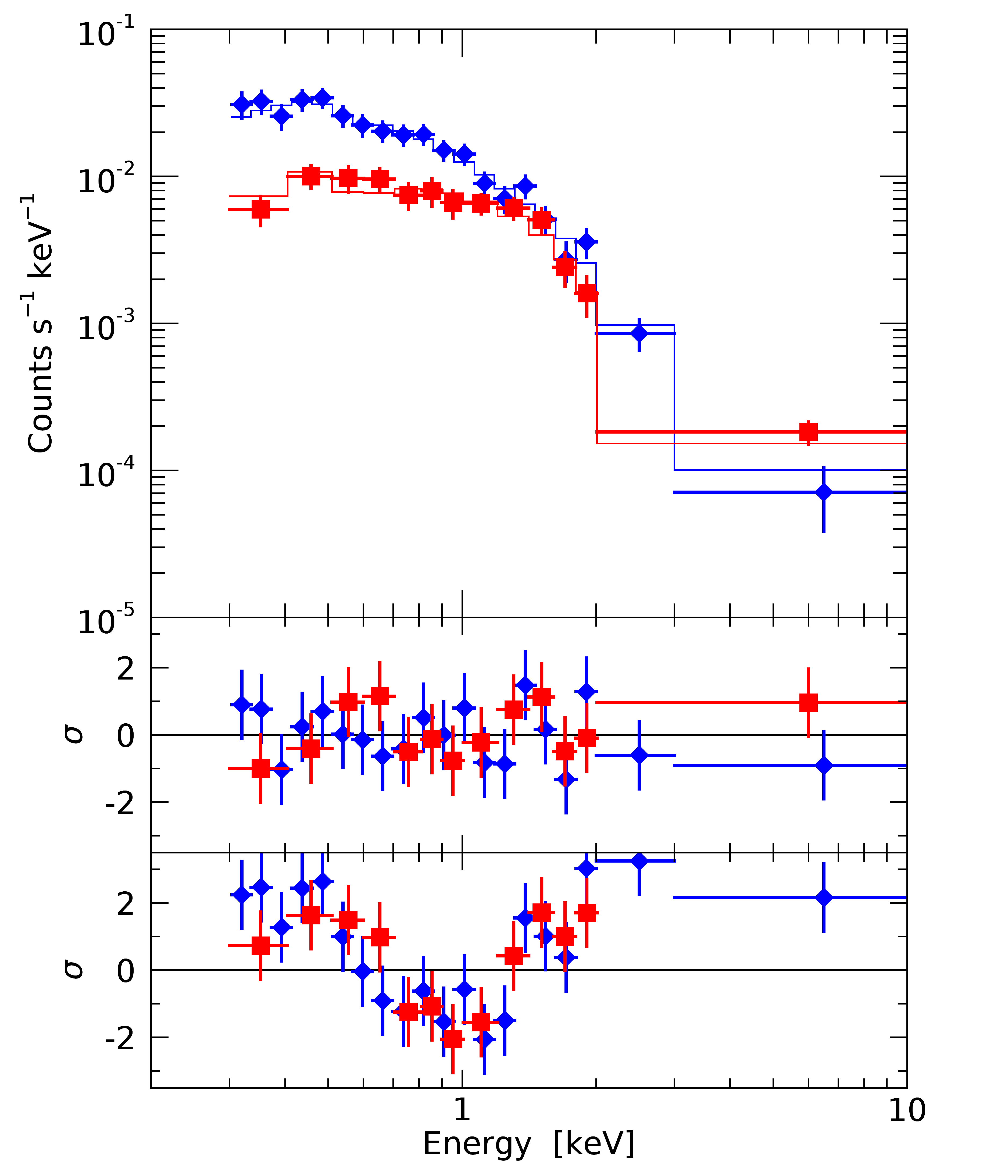}
	\caption{EPIC-pn (blue diamonds) and -MOS (red squares) X-ray spectra of \psra. The best fit power-law is shown in the top panel, and the corresponding residuals in the middle one. The bottom panel shows the residuals obtained by fitting the spectra with an absorbed blackbody. \label{fig:B06fit}}
\end{figure}

\subsection{\psra  (J0630--2834)}

We applied the ML analysis to a circular region of radius $60''$ centred at \coord{06}{30}{51}{01}{-28}{35}{06}{8}. In the energy range \band{0.3}{10}, this yielded     $1021\pm40$ source counts in the pn and $527\pm28$ in the MOS, corresponding to  a total count rate of $(4.0 \pm 0.1) \times 10^{-2}$ cts s$^{-1}$.  

The spectral analysis was carried out in 20 energy bins for the pn and in 13 for the MOS. The two spectra were fitted simultaneously. Among single-component  models, an absorbed power-law gave the best  fit   with $\Gamma=2.85_{-0.15}^{+0.16}$ and  \nh$=(8.5_{-2.1}^{+2.3})\,\times\,10^{20}$~cm$^{-2}$ ($\chi_\nu^2=0.67$ for 30 degrees of freedom (d.o.f.)). 
The error regions of the parameters  are shown in  Fig.~\ref{fig:nh} (a).  A fit with a single blackbody was not acceptable ($\chi_\nu^2 = 3.9$ for 30 d.o.f.).
The best fit power-law spectrum, together with its  residuals and those of the blackbody fit are shown in Fig.~\ref{fig:B06fit}.

The best fit \nh is within $1\sigma$ of the value of $10^{21}$ cm$^{-2}$, derived from the pulsar dispersion measure with the usual assumption of a $10\%$ ionization of the interstellar medium \citep{he13}. Fixing \nh to this value, we get tighter constraints on the photon index, $\Gamma=2.95 \pm 0.06$. The corresponding unabsorbed flux  in the \band{0.2}{10} energy range is $f_{0.2-10} = (1.08 \pm 0.05) \x 10^{-13}$ \flux .

All the above results are in agreement with those of standard spectral analysis of the same data.  In fact, \citet{bec05}  found a power-law with photon index $\Gamma=2.63_{-0.15}^{+0.22}$, while \citet{tep05}, simultaneously fitting the \xmm and \chandra spectra, obtained $\Gamma=3.20_{-0.23}^{+0.26}$ ($1\sigma$ errors in both cases). These authors considered also a power-law plus blackbody model, obtaining acceptable fits with slightly harder power-laws and blackbody temperatures $kT \sim 0.3$~keV.

\begin{figure}
	\centering
		\includegraphics[height=8cm]{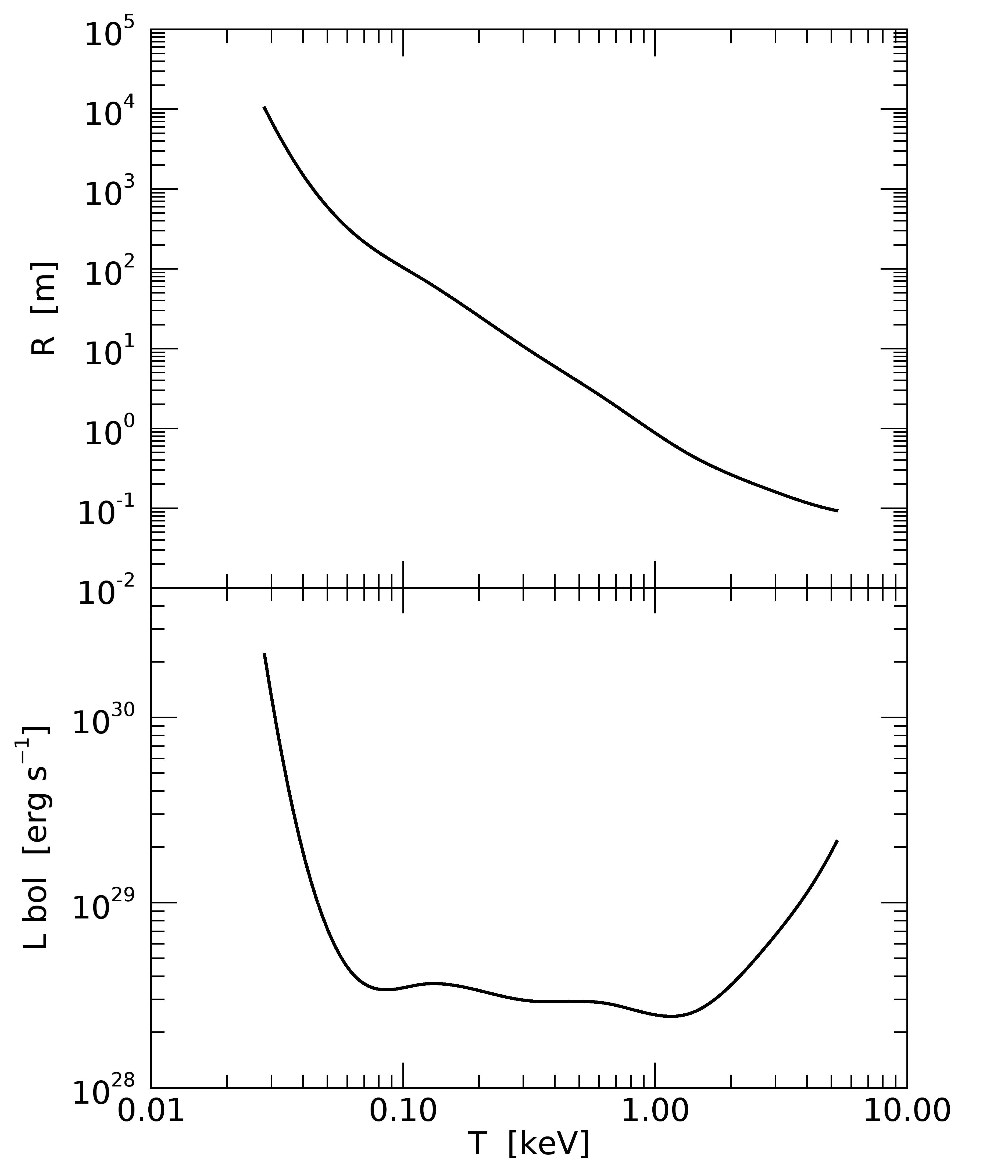}
	\caption{Upper limits ($3\sigma$  c.l.) on the blackbody component in  \psra, as a function of blackbody temperature. The top panel gives the limits on the emitting radius, while the lower panel gives the corresponding limits on the bolometric luminosity, computed as $L_{\rm bol} = \pi R^2 \sigma T^4$. \label{fig:B06upplum}}
\end{figure}

Using a power-law plus blackbody model we found that several combinations of blackbody temperature and normalization, which give only a limited flux contribution, are consistent with the data. The best fit parameters were $\Gamma=2.92\pm0.07$ and $kT\sim0.02$~keV, while the emitting radius is poorly constrained. However, the addition of a blackbody is not statistically required. Therefore, we  derived an upper limit on its intensity as follows. We fitted the spectrum with a blackbody plus power-law model (\nh fixed to $10^{21}$ cm$^{-2}$) and computed the $3\sigma$  confidence ranges of the blackbody normalization and temperature, leaving the power-law parameters free to vary. The results are shown in Fig.~\ref{fig:B06upplum}, where the upper panel refers to  the  emitting radius and the lower panel to the  bolometric luminosity. The latter has been computed as 
$L_{\rm bol}= \pi R^2 \sigma T^4$, because we are considering emission from a  hot spot on the stellar surface.

\begin{figure*}
	\centering
	\includegraphics[width=7.cm]{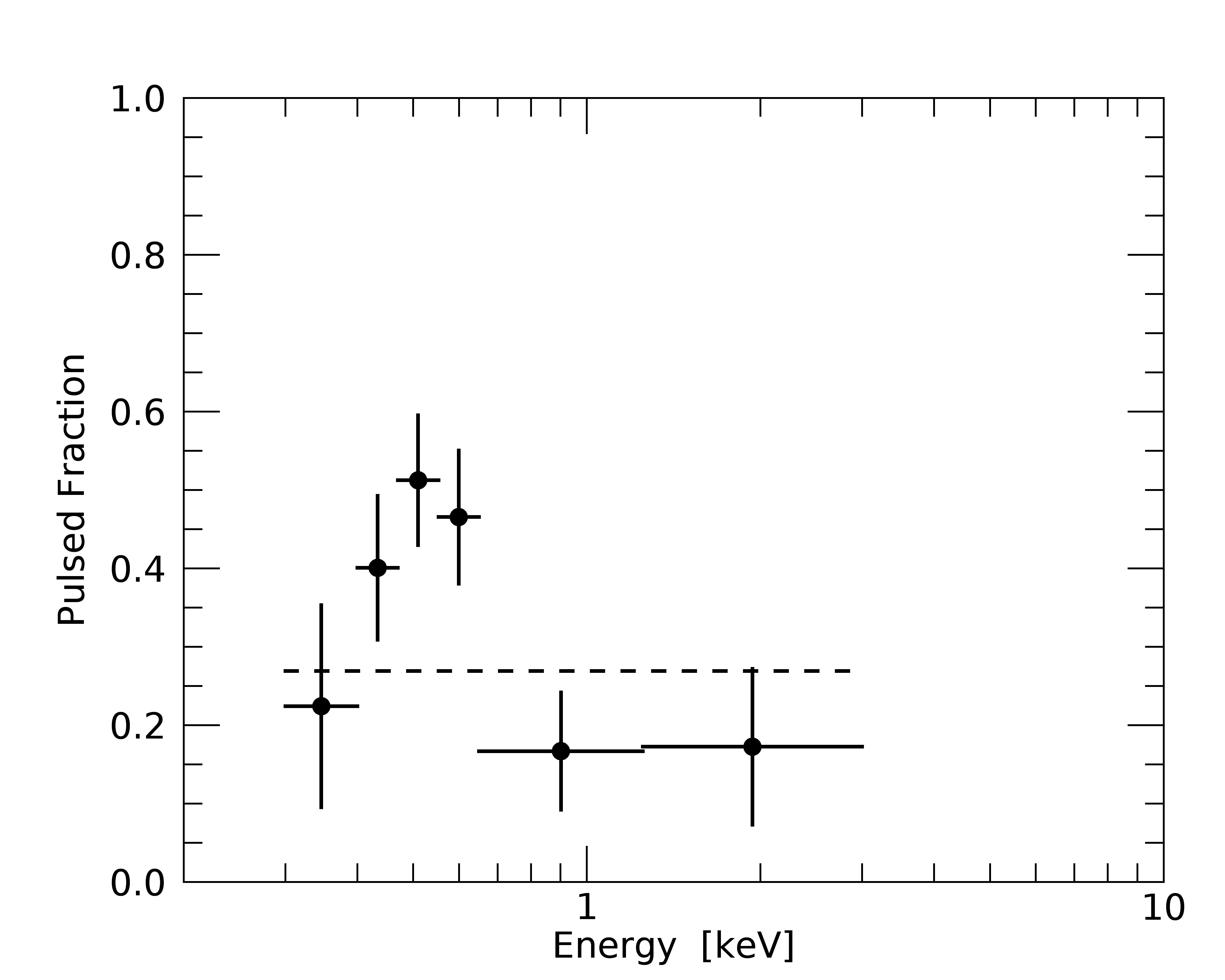}
	\includegraphics[width=7.cm]{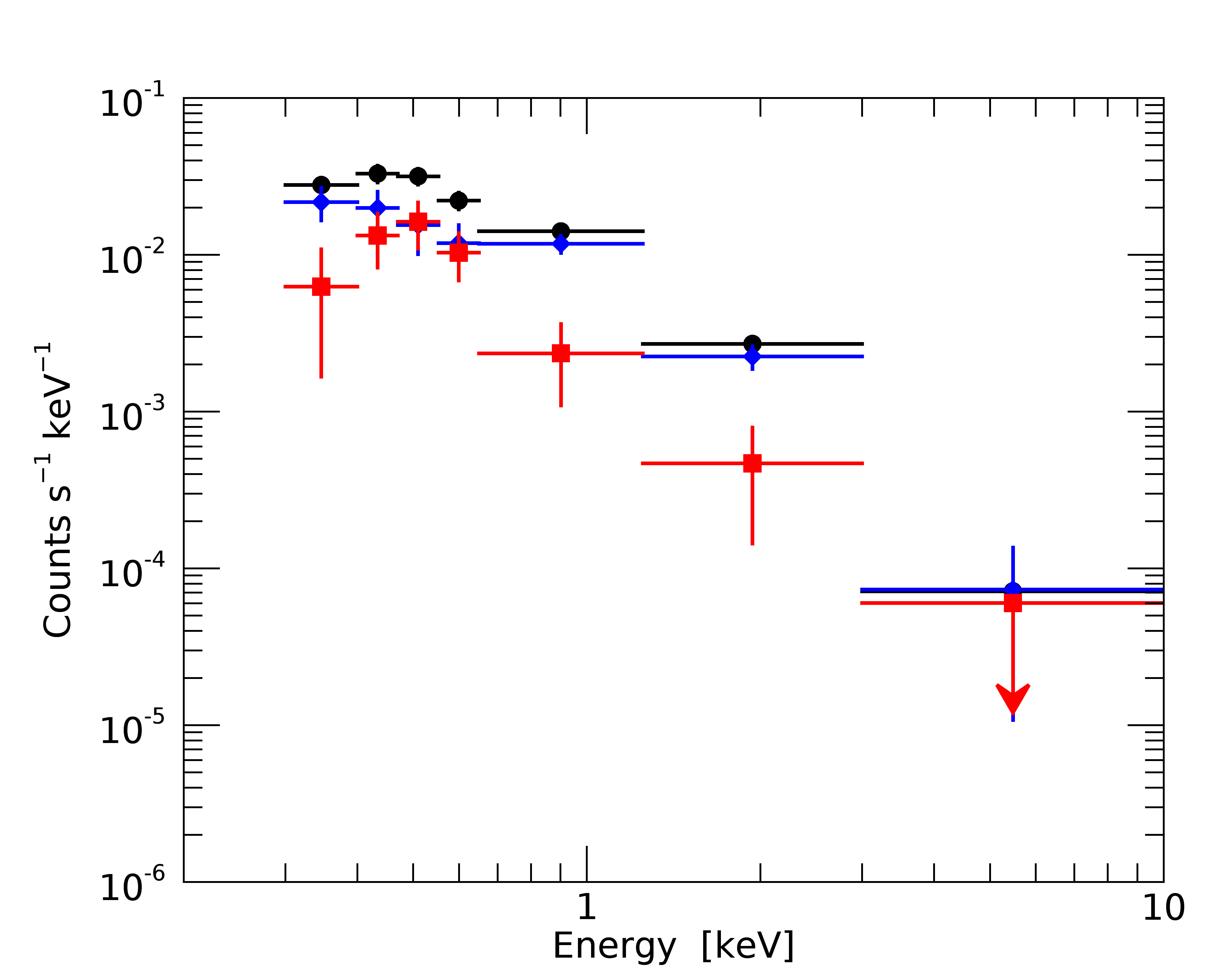}
	\caption{Left: pulsed fraction of \psra as a function of energy, obtained with  the 3D-ML analysis. 
	Right: EPIC-pn total (black circles),  unpulsed (blue diamonds),  and pulsed (red squares) spectra of \psra. \label{fig:B063D}}	
\end{figure*}

To investigate the spectral properties of the pulsed and unpulsed emission, 
we performed a 3D-ML analysis on the pn data in 7 energy bins in the range \band{0.3}{10}. 
The phases of the detected counts were computed using the period at the epoch of the \xmm observation, $P=1.2444225976(2)$~s,  obtained from the ephemeris reported in the   online\footnote{http://www.atnf.csiro.au/people/pulsar/psrcat/} ATNF Pulsar  Catalogue \citep{man05}.
The pulsed fraction is higher in the  \band{0.40}{0.65} range, with an average value of $0.46\pm0.05$ and decreases at lower and higher energies (Fig.~\ref{fig:B063D} left). 
The energy-dependence of the pulsed fraction is reflected also in the different shape of the pulsed and unpulsed spectra (Fig.~\ref{fig:B063D} right). 
The unpulsed spectrum is well fitted by a power-law with the same slope of that of the  total spectrum, while that
of the pulsed emission requires a significantly steeper power-law or a blackbody (Table~\ref{tab:B06par}).

\setlength{\tabcolsep}{1em}
\begin{table*}[htbp!]
\centering \caption{Spectral parameters of \psra}
\label{tab:B06par}

\begin{tabular}{lcccccc}
\toprule 
Model	& \nh	& $\Gamma$		& PL Norm		& $kT$	& $R_{\rm BB}\,^{\rm b}$	& $\chi_{\nu}^2$/d.o.f.	\\[5pt]
		& $10^{20}$~cm$^{-2}$ &	& $^{\rm a}$ 	& keV	& m					& 			\\[5pt]

\midrule\\[-5pt]

PL		& $8.5_{-2.1}^{+2.3}$	& $2.85_{-0.15}^{+0.16}$	& $1.4 \pm 0.1$	& \dots	& \dots	& 0.67/30	\\[5pt]
PL		& $10\,^{\rm c}$	& $2.95 \pm 0.06$	& $1.44 \pm 0.05$	& \dots	& \dots	& 0.66/31	\\[5pt]
PL+BB	& $10\,^{\rm c}$	& $2.92 \pm 0.07$	& $1.42 \pm 0.05$	&$\sim\!0.02$	& $<4.5\x10^5$	& 0.67/29	\\[5pt]		

\midrule\\[-5pt]

PL unpulsed$\,^{\rm d}$	& $10\,^{\rm c}$	& $2.7 \pm 0.2$	& $1.1 \pm 0.1$	& \dots	& \dots	& 0.64/5	\\[5pt]
PL pulsed$\,^{\rm d}$	& $10\,^{\rm c}$	& $3.8_{-0.5}^{+0.6}$	& $0.3 \pm 0.1$	& \dots	& \dots	& 0.93/5	\\[5pt]
BB pulsed$\,^{\rm d}$	& $10\,^{\rm c}$	& \dots	& \dots	& $0.10_{-0.01}^{+0.02}$	& $150_{-50}^{+80}$	& 0.75/5	\\[5pt]

\bottomrule\\[-5pt]
\end{tabular}

\raggedright
\scriptsize
Joint fits of pn + MOS spectra. Errors at $1\sigma$. \\
$^{\rm a}$ Normalization of the power-law at 1 keV in units of $10^{-5}$~photons~cm$^{-2}$~s$^{-1}$~keV$^{-1}$. \\
$^{\rm b}$ With an assumed distance of 0.32 kpc.\\ 
$^{\rm c}$ Fixed value.\\
$^{\rm d}$ pn spectrum only.\\
\end{table*}

\subsection{\psrb  (J0922+0638)} 

\begin{figure}
	\centering
	\includegraphics[height=8cm]{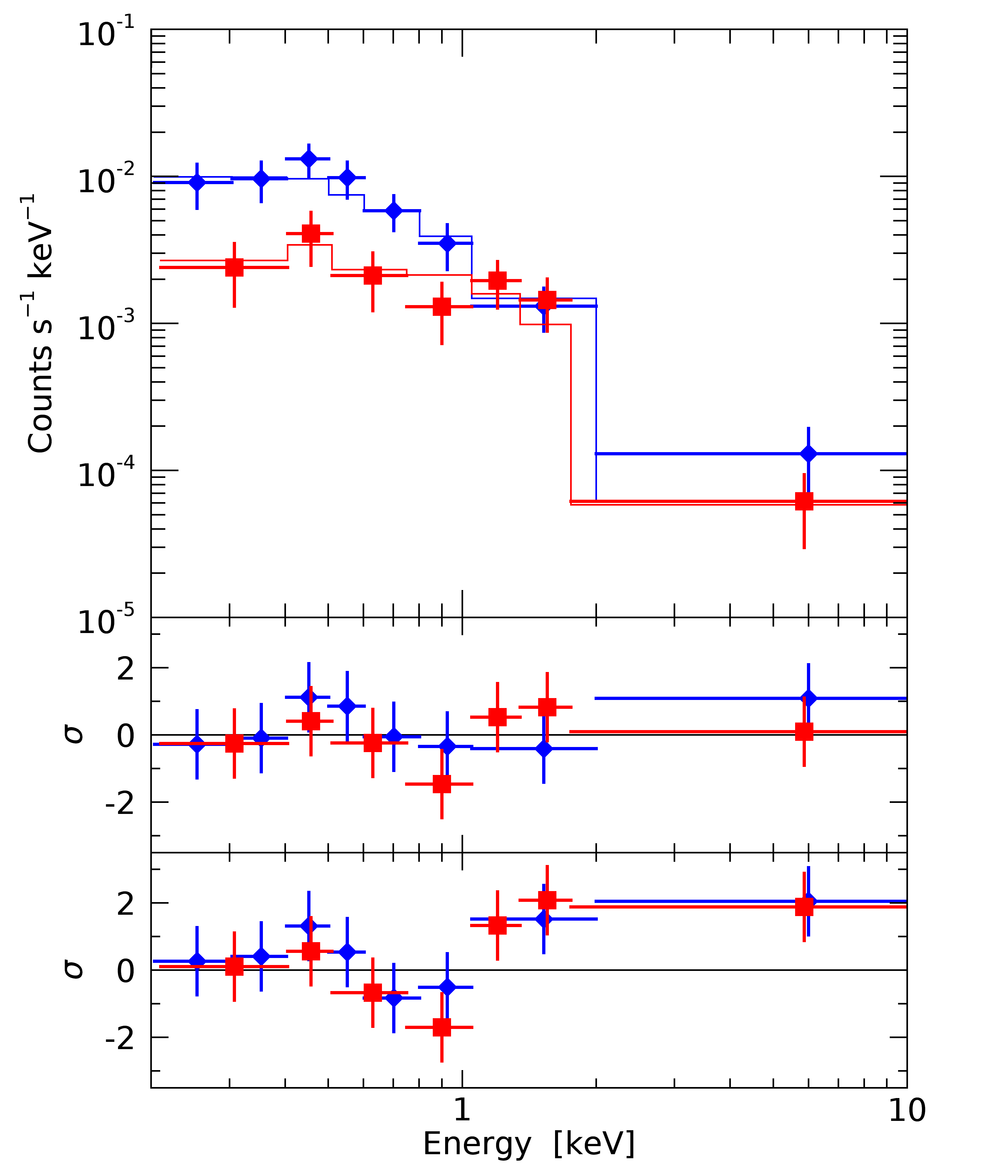}
	\caption{EPIC pn (blue diamonds) and MOS (red squares) X-ray spectra of \psrb. The best fit power-law is shown in the top panel, and the corresponding residuals in the middle one. The bottom panel shows the residuals obtained by fitting the spectra with an absorbed blackbody. \label{fig:B09fit}}
\end{figure}

The faint X-ray counterpart of \psrb was first reported by \citet{pri15}. Based on a spectrum  with $171\pm18$  counts, they rejected the blackbody model and found a good power-law fit with $\Gamma=2.3_{-0.4}^{+0.8}$. 
Our ML analysis of  the same data  was carried out in circular region  with  radius of $45''$ 
centred at \coord{09}{22}{14}{20}{+06}{38}{33}{7}. In the energy range \band{0.2}{10} we obtained $223\pm25$ (pn) and $91\pm15$ (MOS) source  counts, corresponding to a total count rate of $(1.3 \pm 0.1) \times 10^{-2}$ cts s$^{-1}$. 

We simultaneously fitted the pn (8 energy bins) and the MOS (7 energy bins) spectra. The single-component model that best fits the data is an absorbed power-law (see Fig.~\ref{fig:nh} (b) and Table~\ref{tab:B09par}). The single blackbody model is statistically unacceptable ($\chi_\nu^2=1.92$ for 12 d.o.f.).
We show the comparison between the power-law and the blackbody best fits in Fig.~\ref{fig:B09fit}.
Fixing the absorption to the DM-based value $8 \x 10^{20}$ cm$^{-2}$,  we obtain $\Gamma=3.1\pm0.2$ and an unabsorbed flux $f_{0.2-10} = (3.0 \pm 0.4) \x 10^{-14}$ \flux. 

\begin{figure}
	\centering
		\includegraphics[height=8cm]{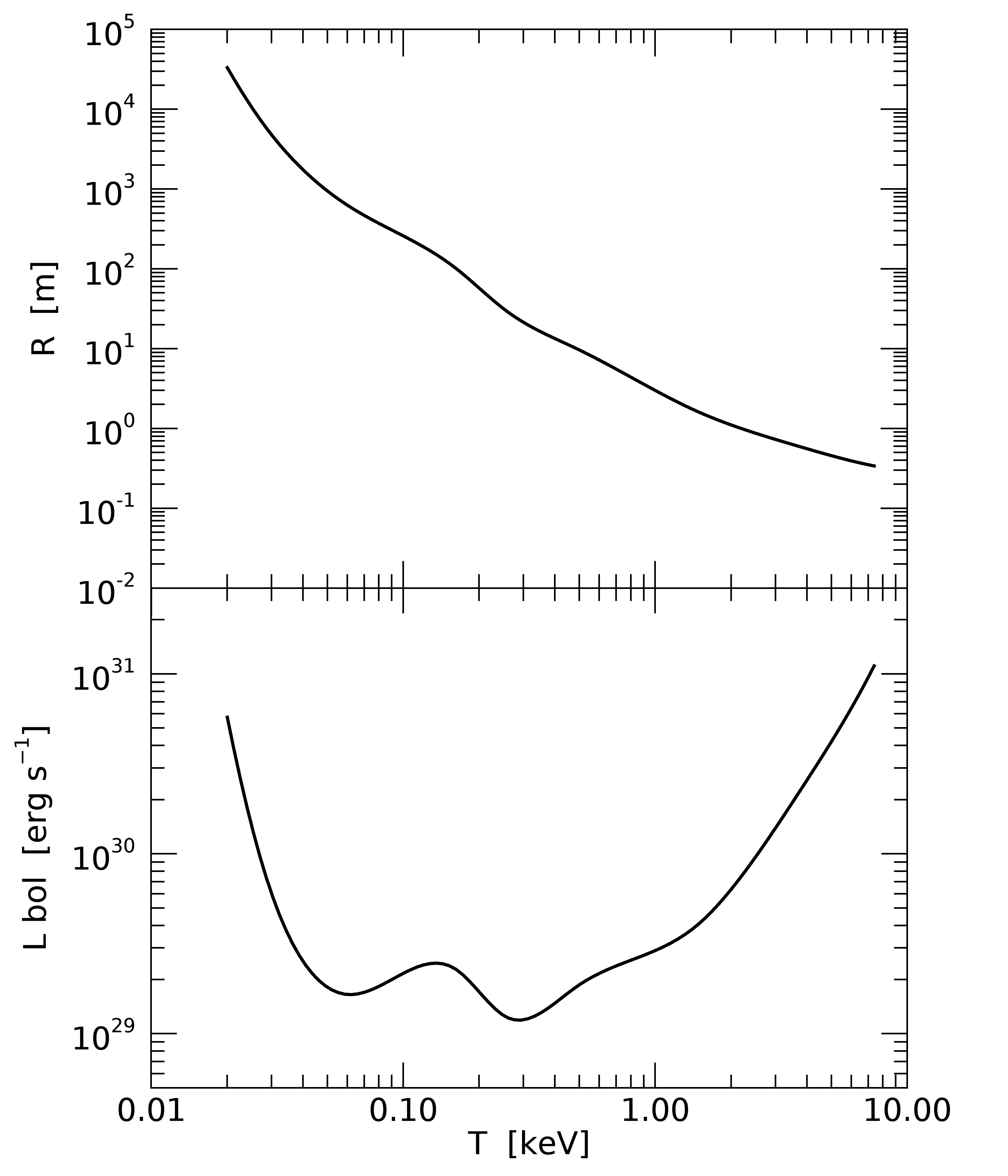}
	\caption{Upper limits ($3\sigma$  c.l.) on the blackbody component in  \psrb, as a function of blackbody temperature. The top panel gives the limits on the emitting radius, while the lower panel gives the corresponding limits on the bolometric luminosity, computed as $L_{\rm bol} = \pi R^2 \sigma T^4$. \label{fig:B09upplum}}
\end{figure}

Although a good fit is also obtained with a  power-law  plus blackbody (Table~\ref{tab:B09par}), the addition of a blackbody is not statistically required. 
Therefore, as we did for \psra, we derived the upper limits on its intensity as a function of $T$ shown in  Fig.~\ref{fig:B09upplum}. 

\setlength{\tabcolsep}{1em}
\begin{table*}[htbp!]
\centering \caption{Spectral parameters of \psrb}
\label{tab:B09par}
\begin{tabular}{lcccccc}
\toprule 
Model	& \nh	& $\Gamma$		& PL Norm	& $kT$	& $R_{\rm BB}\,^{\rm b}$	& $\chi_{\nu}^2$/d.o.f.	\\[5pt]
		& $10^{20}$~cm$^{-2}$ &	& $^{\rm a}$ 	& keV	& m			& 			\\[5pt]

\midrule\\[-5pt]

PL		& $4_{-3}^{+4}$	& $2.7_{-0.4}^{+0.5}$	& $3.1_{-0.5}^{+0.6}$	& \dots	& \dots	& 0.58/12	\\[5pt]
PL		& $8\,^{\rm c}$	& $3.1 \pm 0.2$	& $3.6 \pm 0.4$	& \dots	& \dots	& 0.61/13	\\[5pt]
PL+BB	& $8\,^{\rm c}$	& $2.2 \pm 0.6$	& $2.7_{-1.0}^{+0.9}$	& $0.11_{-0.03}^{+0.02}$	& $214_{-67}^{+125}$	& 0.48/11\\[5pt]		

\bottomrule\\[-5pt]
\end{tabular}

\raggedright
\scriptsize
Joint fits of pn + MOS spectra. Errors at $1\sigma$. \\
$^{\rm a}$ Normalization of the power-law at 1 keV in units of $10^{-6}$~photons~cm$^{-2}$~s$^{-1}$~keV$^{-1}$. \\
$^{\rm b}$ With an assumed distance of 1.1 kpc.\\ 
$^{\rm c}$ Fixed value.\\ 
\end{table*}

\subsection{\psrc  (J0117+5914)}

This faint pulsar,  observed with \xmm\ only  for less than 6 ks, is the one with the smallest number of counts in our sample. 
\citet{pri15} obtained a poorly constrained  spectrum, which could be fitted equally well by a power-law with $\Gamma=3.3\pm0.5$ or by a blackbody with temperature $kT = 0.18 \pm 0.03$ keV and emitting radius $R=322\pm161$ m (for $d=1.77$ kpc).

We applied the ML analysis in a circular region centred at \coord{01}{17}{38}{24}{+59}{15}{07}{8} with a radius of $60''$, from which we excluded a circle centred in \coord{01}{17}{34}{87}{+59}{15}{15}{4} with a radius of $20''$ to avoid a nearby source.
This resulted in  $85 \pm 13$ pn counts and $44 \pm 9$ MOS counts in the range \band{0.2}{10},  corresponding to a total count rate of $(2.2 \pm 0.3) \times 10^{-2}$ cts s$^{-1}$.

Simultaneously fitting the pn and the MOS spectra it was impossible to discriminate between a power-law  ($\chi_\nu^2=0.58$ for 7 d.o.f.) and a blackbody ($\chi_\nu^2=0.62$ for 7 d.o.f.). However, restricting the analysis only to the pn spectrum, which is of better statistical quality, and fixing the absorption to the DM-based value of $1.5 \x 10^{21}$ cm$^{-2}$, we found that the blackbody is preferred ($\chi_\nu^2=0.32$ for 4 d.o.f., compared to $\chi_\nu^2=1.80$ for the power-law).  The best fit  temperature is $kT = 0.17 \pm 0.02$ keV, the   radius of the emitting area  is $R=405_{-90}^{+110}$ m, and the \band{0.2}{10} unabsorbed flux is $f_{0.2-10} = (4.7 \pm 0.7) \x 10^{-14}$ \flux 
(see Table~\ref{tab:B01par} and Fig.~\ref{fig:nh} (c)).
Letting the absorption free to vary, also the power-law is acceptable, but with an unphysical photon index $\Gamma = 6.5_{-2.0}^{+3.0}$.

\setlength{\tabcolsep}{1em}
\begin{table*}[htbp!]
\centering \caption{Spectral parameters of \psrc}
\label{tab:B01par}

\begin{tabular}{lcccccc}
\toprule 
Model	& \nh	& $\Gamma$		& PL Norm	& $kT$	& $R_{\rm BB}\,^{\rm b}$	& $\chi_{\nu}^2$/d.o.f.	\\[5pt]
		& $10^{20}$~cm$^{-2}$ &	& $^{\rm a}$ 	& keV	& m			& 			\\[5pt]

\midrule\\[-5pt]

PL		& $50_{-20}^{+30}$	& $5.5_{-1.3}^{+2.0}$	& $4_{-2}^{+5}$	& \dots	& \dots	& 0.58/7	\\[5pt]
BB		& $8_{-7}^{+12}$	& \dots	& \dots	& $0.20 \pm 0.04$	& $<566$	& 0.62/7	\\[5pt]		
BB		& $15\,^{\rm c}$	& \dots	& \dots	& $0.17 \pm 0.02$	& $405_{-90}^{+110}$	& 0.60/8	\\[5pt]		

\bottomrule\\[-5pt]
\end{tabular}

\raggedright
\scriptsize
Joint fits of pn + MOS spectra. Errors at $1\sigma$. \\
$^{\rm a}$ Normalization of the power-law at 1 keV in units of $10^{-5}$~photons~cm$^{-2}$~s$^{-1}$~keV$^{-1}$. \\
$^{\rm b}$ With an assumed distance of 1.77 kpc.\\ 
$^{\rm c}$ Fixed value.\\ 
\end{table*}

\subsection{\psrd  (J1136+1551)}

\psrd has both optical \citep{zha08,zha13} and X-ray \citep{kar06} counterparts. According to these works, its  X-ray spectrum   can be described by either a   power-law or a   blackbody (or a combination of them), but the statistics is  too poor to distinguish between these models.  

\psrd  was observed with five separate pointings spanning about one month. With a simultaneous fit of the separate spectra,  \citet{sza17} found a power-law   with $\Gamma \approx 2.5$ and reported that the addition of a blackbody  with $T=0.25_{-0.03}^{+0.05}$~keV and a radius of $14_{-5}^{+7}$~m (for $d = 350$ pc) is also consistent with the data.

We verified that the analysis of the single observations gave fully consistent results. Therefore, we report only those  obtained by combining  the five pointings  to extract  a single pn and a single MOS spectrum. 
We applied the ML analysis in a circular region of radius $35''$ and centred at \coord{11}{36}{02}{527}{15}{50}{59}{90}. We detected the source with a count rate of $(1.06 \pm 0.05) \times 10^{-2}$ cts s$^{-1}$ in the energy range \band{0.2}{10}. For the spectral analysis the $887 \pm 46$ pn and $363 \pm 29$ MOS counts  were divided into 19 and 11 energy bins, respectively.

The pn and the MOS spectra were fitted simultaneously. 
We could not obtain acceptable fits with a single power-law or  blackbody, or any combination of them (including a power-law plus two blackbodies), because, in all cases, significant residuals were  present below 0.5 keV.
It is very unlikely  that this is due to an instrumental or calibration problem  because the spectra of two nearby  soft and very bright sources do not show similar features.

Excluding the energy channels below 0.5 keV, we could obtain a good fit with an absorbed power-law with photon index $\Gamma = 3.2\pm 0.4$ and \nh$= (8_{-6}^{+7}) \x 10^{20}$~cm$^{-2}$  (Fig.~\ref{fig:nh}, panel (d)). With \nh fixed to the DM-based value of $1.5 \x 10^{20}$ cm$^{-2}$ we obtained a photon index of $2.8 \pm 0.1$ ($\chi_\nu^2=0.82$, for 18 d.o.f.). A blackbody fit was rejected ($\chi_\nu^2 \approx 2$ both for free and fixed \nh).

\begin{figure}
	\centering
	\includegraphics[width=7cm]{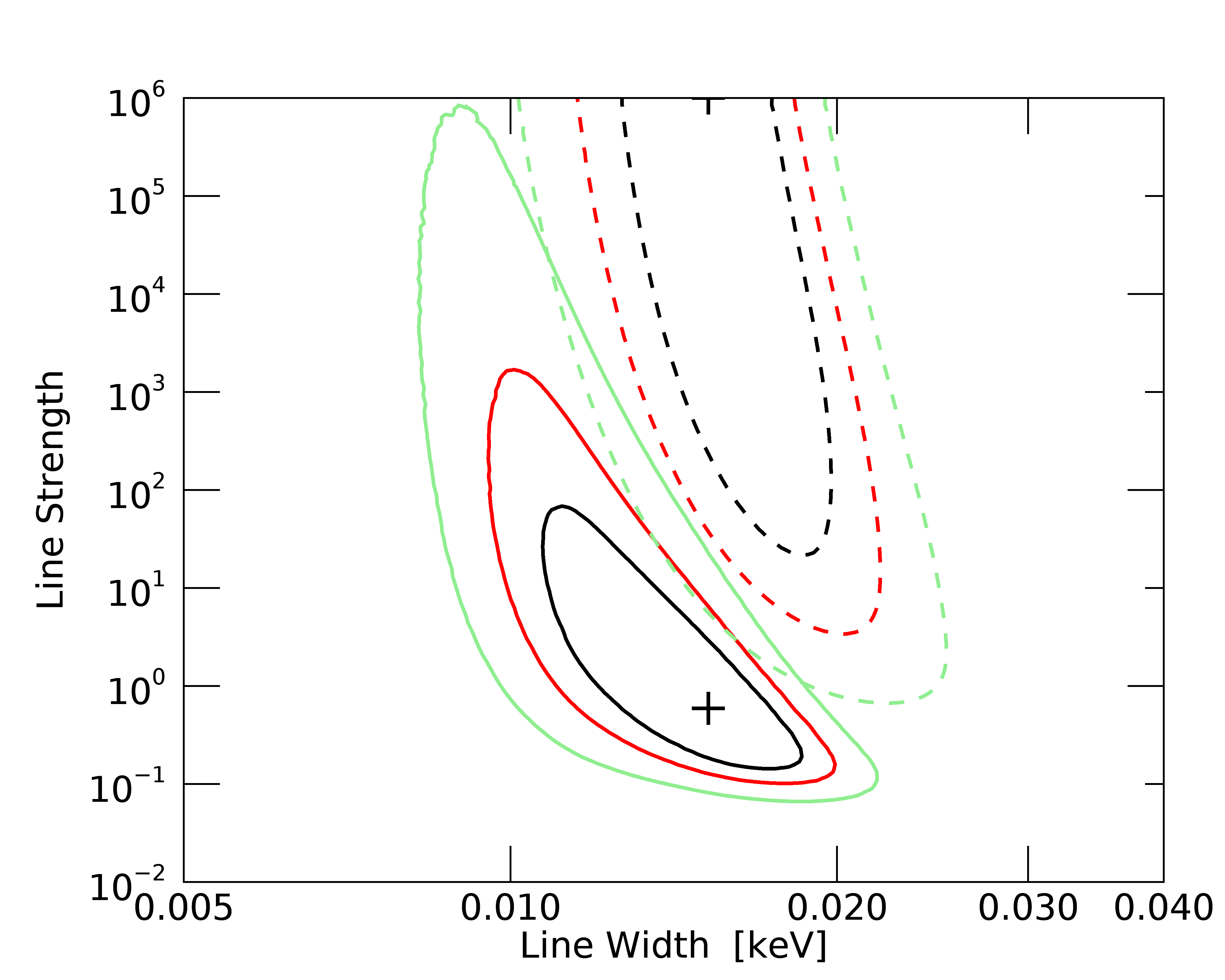}
	\caption{Confidence regions (1, 2 and 3 $\sigma$) of the best fit line strength versus  width of the two absorption lines in the spectrum of \psrd  (dashed lines: $E_1\sim0.22$ keV, solid lines: $E_2\sim0.44$ keV). The best fit values are represented by a black plus. 
	\label{fig:B11con}}
\end{figure}

\begin{figure}
	\centering
	\includegraphics[height=8cm]{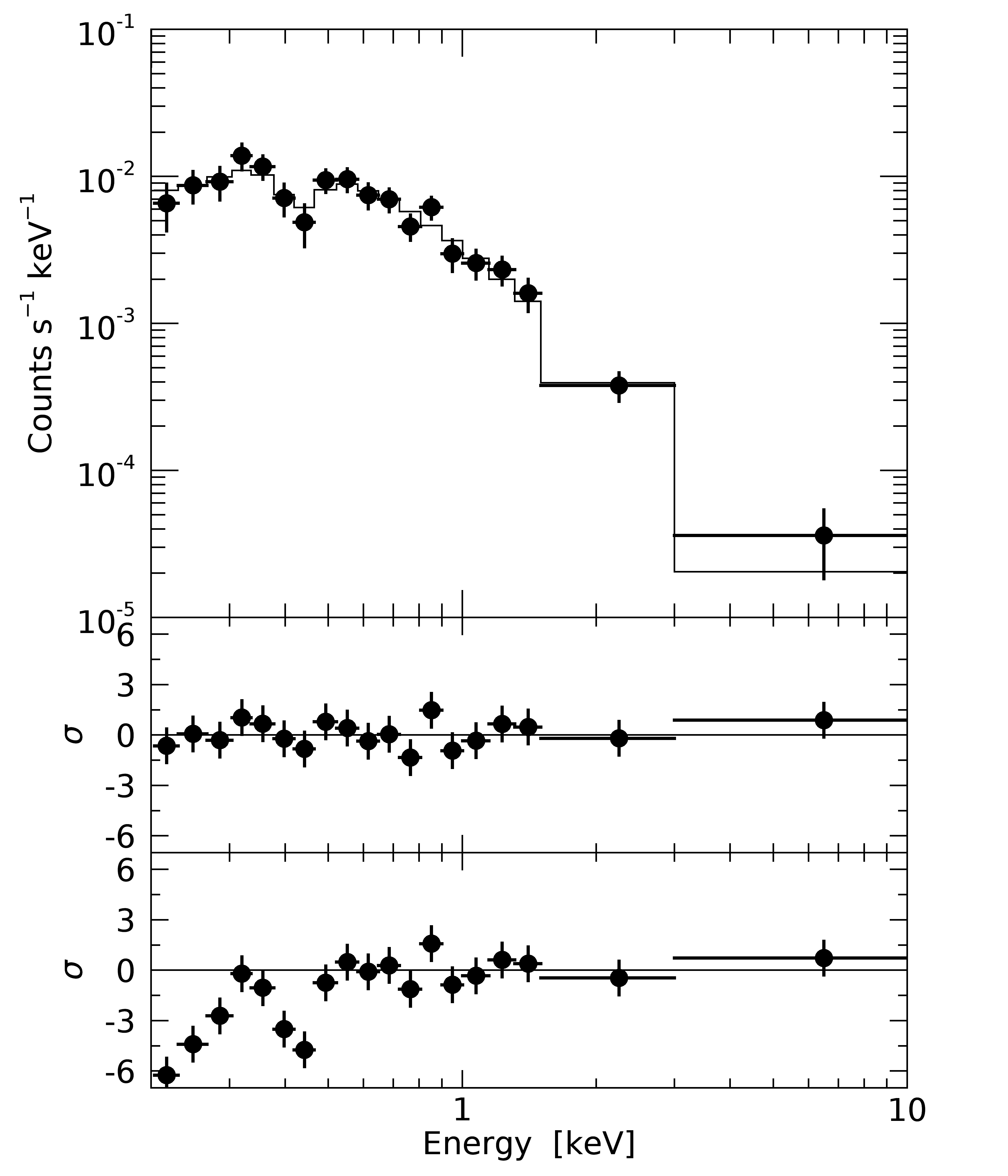}
	\caption{EPIC-pn spectrum of \psrd . The top panel shows the best fit with an absorbed power-law plus two absorption lines at $E_1\sim 0.22$ and $E_2\sim 0.44$ keV. The corresponding residuals are shown in the middle panel. The bottom panel shows the residuals obtained by the best fit with a single power-law in the range \band{0.5}{10}. \label{fig:B11fit}}
\end{figure}

A power-law fit of the  pn spectrum over the whole energy range suggests the presence of two absorption lines  at $\sim\!0.2$ keV and $\sim\!0.44$~keV. We therefore fitted it with a  model consisting of a  power-law with \nh fixed to $1.5 \times 10^{20}$ cm$^{-2}$  and two gaussian absorption lines. We constrained the lines to be centred at $E_2=2\times E_1$ and to have the same width  $\sigma_1=\sigma_2$.
This model gave  a good fit with $E_1=0.222\pm0.006$~keV and  $\sigma_1=0.015_{-0.004}^{+0.012}$~keV ($\chi_\nu^2 = 0.78$ for 13 d.o.f.). The best fit photon index is $2.9 \pm 0.2$,   in good agreement with that found in the hardest part of the spectrum.
The strength of the lines are poorly constrained, but they are $3 \sigma$ above the 0 level (see Fig.~\ref{fig:B11con}).
The best fit spectrum, together with its  residuals and those of the single power-law fit are shown in Fig.~\ref{fig:B11fit}.

A good fit could also be obtained with a power-law plus blackbody with an absorption line at $\sim\!0.45$~keV ($\chi_\nu^2= 0.83$ for 11 d.o.f.). However, the \nh$=(22_{-10}^{+12}) \x 10^{20}$~cm$^{-2}$ is much higher than expected, the power-law is quite steep ($\Gamma=3.5_{-0.5}^{+0.7}$), and the blackbody has $kT=0.045\pm0.010$~keV and a poorly constrained radius ($1<R<70$~km). This is compatible with emission from the cooling of the whole surface, but for an old pulsar such as \psrd we would expect a lower temperature, unless some reheating mechanism is operating.

To assess  the statistical significance of the lines we estimated the probability of obtaining by chance a fit improvement as the observed one  through Monte Carlo simulations.
We simulated  pn spectra of \psrd  with the same  exposure time as in our observations  using a model without lines, i.e.  the best fit power-law model with $\Gamma=2.36$ (Table~\ref{tab:B11par}). 
We fitted each simulated spectrum with a single power-law and with a power-law plus two harmonically spaced lines and computed the ratio of  the corresponding $\chi^2$ values $F$ = $\chi_{\rm PL}^2$ / $\chi_{\rm PL+lines}^2$.
We found a probability of $8 \times 10^{-4}$ to have a  fit improvement better than that observed in the real data (i.e. $F>2.47$), corresponding to a $\sim\!3.1\sigma$ significance of the lines.

\begin{figure*}
	\centering
	\includegraphics[width=7.cm]{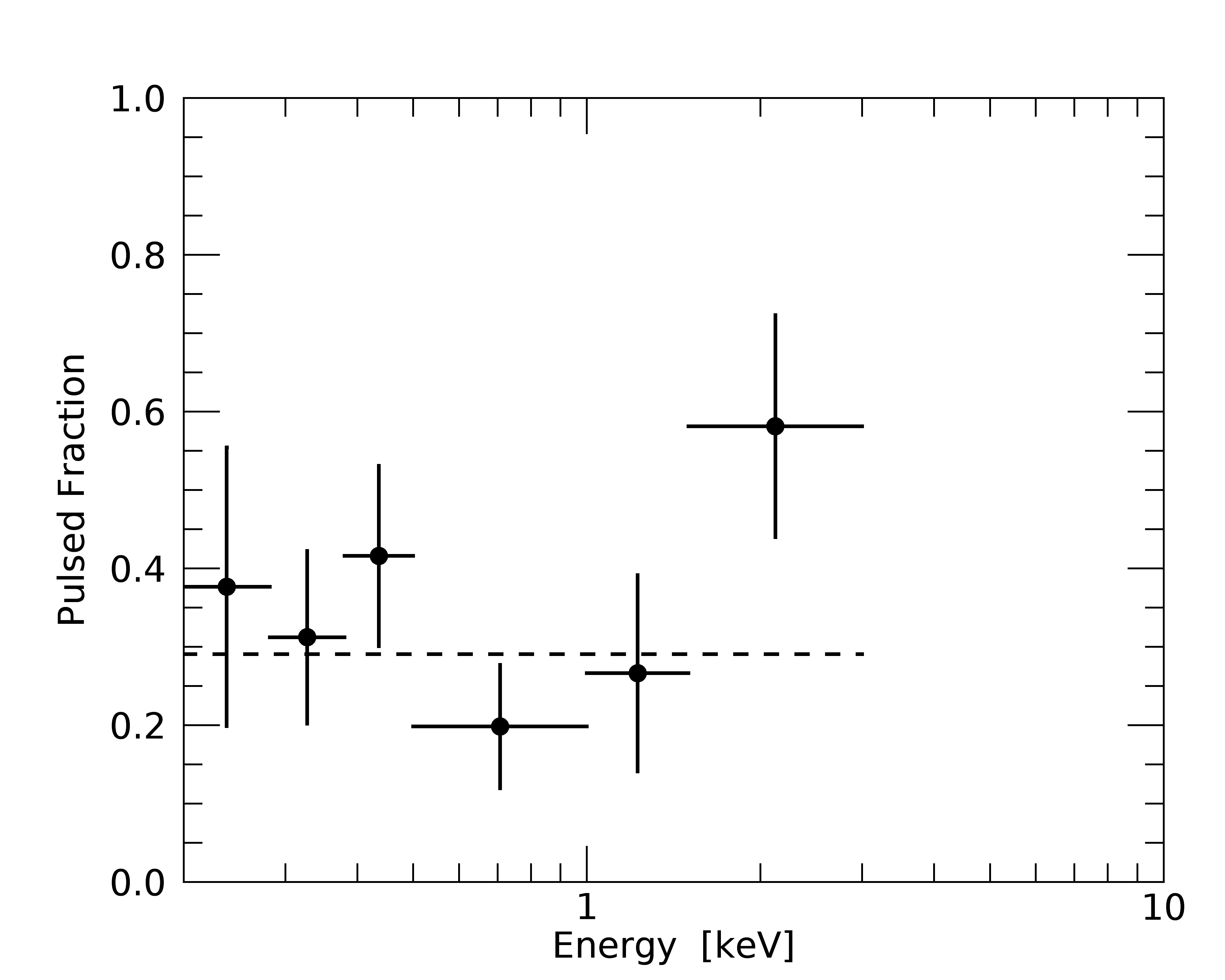}
	\includegraphics[width=7.cm]{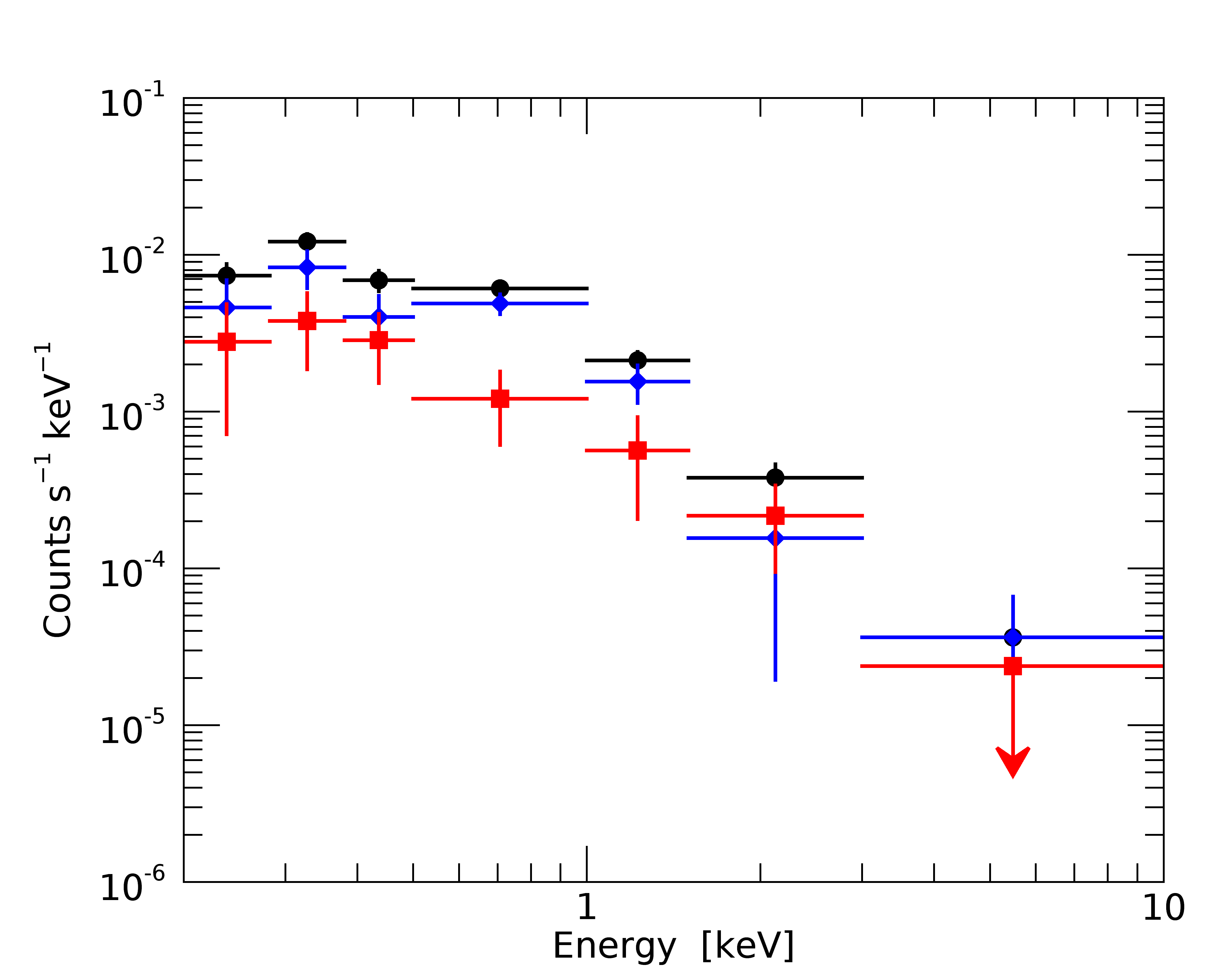}
	\caption{Left: pulsed fraction of \psrd as a function of energy, obtained with  the 3D-ML analysis. 
	Right: EPIC-pn total (black circles),  unpulsed (blue diamonds),  and pulsed (red squares) spectra of \psrd. \label{fig:B113D}}	
\end{figure*}

For the 3D-ML spectral-timing analysis, the phases were computed using the period at the epoch of the \xmm observation, $P=1.187916418694(5)$~s given by \citet{sza17}. 
We found an average pulsed fraction of $0.29\pm0.05$, and no significant energy  variation  over the  range \band{0.2}{3} in the pn (see Fig.~\ref{fig:B113D}, left). A hint of pulsation was also found in the MOS, with a pulsed fraction of $0.25\pm0.07$ in the range \band{0.2}{3}, but it was no possible to extract the 3D spectra.
The pn spectra of the total, pulsed and unpulsed flux are plotted in Fig.~\ref{fig:B113D} (right).
The unpulsed spectrum cannot be fitted by a single power-law ($\chi_\nu^2 = 2.12$ for 5 d.o.f.), while  it is well fitted by the same absorption features of the total model. On the contrary, the spectrum of the pulsed emission does not require the addition of the absorption lines, and a power-law with $\Gamma = 2.4_{-0.4}^{+0.5}$  satisfactorily fits the spectrum.
All the results of the spectral fits are summarized in Table~\ref{tab:B11par}.

\setlength{\tabcolsep}{0.25em}
\begin{table*}[htbp!]
\centering \caption{Spectral parameters of \psrd}
\label{tab:B11par}

\begin{tabular}{lcccccccc}
\toprule 
Model	& \nh	& $\Gamma$	& PL Norm	& $kT$	& $R_{\rm BB}\,^{\rm b}$	& $E_1$	& $\sigma_1$	& $\chi_{\nu}^2$/d.o.f.	\\[5pt]
		& $10^{20}$~cm$^{-2}$ &	& $^{\rm a}$& keV	& km		& keV	& keV			&	 			\\[5pt]

\midrule\\[-5pt]

PL ($E > 0.5$ keV)  	& $8_{-6}^{+7}$	& $3.2 \pm 0.4$	& $3.6_{-0.7}^{+0.9}$	& \dots	& \dots	& \dots	& \dots	& 0.81/17	\\[5pt]
PL ($E > 0.5$ keV) 	& $1.5\,^{\rm c}$	& $2.8 \pm 0.1$	& $2.9 \pm 0.1$	& \dots	& \dots	& \dots	& \dots	& 0.82/18	\\[5pt]

PL$\,^{\rm d}$ 	& $1.5\,^{\rm c}$	& $2.36 \pm 0.08$	& $2.60 \pm 0.15$	& \dots	& \dots	& \dots	& \dots	& 1.925/17	\\[5pt]

PL+2L$\,^{\rm d}$		& $<2.5$	& $2.8 \pm 0.2$	& $2.7_{-0.2}^{+0.3}$	& \dots	& \dots	& $0.221\pm0.006$	& $0.016_{-0.004}^{+0.011}$	& 0.80/12	\\[5pt]	
PL+2L$\,^{\rm d}$		& $1.5\,^{\rm c}$	& $2.9 \pm 0.2$	& $2.9 \pm 0.2$	& \dots	& \dots	& $0.222\pm0.006$	& $0.015_{-0.004}^{+0.012}$	& 0.78/13	\\[5pt]
PL+2L & $1.5\,^{\rm c}$	& $2.9 \pm 0.1$	& $2.90 \pm 0.15$	& \dots	& \dots	& $0.225\pm0.006$	& $0.017_{-0.001}^{+0.020}$	& 1.10/24	\\[5pt]

PL+BB+1L$\,^{\rm d}$	& $22_{-10}^{+12}$	& $3.5_{-0.5}^{+0.7}$	& $5_{-1}^{+2}$	& $0.045\pm0.010$	& $<70$	& $0.450_{-0.008}^{+0.007}$	& $0.010_{-0.001}^{+0.020}$	& 0.83/11	\\[5pt] 
PL+2L unpulsed$\,^{\rm d}$	& $1.5\,^{\rm c}$	& $2.9 \pm 0.2$			& $2.1\pm0.3$	& \dots	& \dots	& $0.222\,^{\rm c}$	& $0.015\,^{\rm c}$	& 0.58/5	\\[5pt]
PL+2L pulsed$\,^{\rm d}$	& $1.5\,^{\rm c}$	& $3.0_{-0.6}^{+0.8}$	& $0.8\pm0.3$	& \dots	& \dots	& $0.222\,^{\rm c}$	& $0.015\,^{\rm c}$	& 0.44/5	\\[5pt]
PL pulsed$\,^{\rm d}$		& $1.5\,^{\rm c}$	& $2.4_{-0.4}^{+0.5}$	& $0.8\pm0.2$	& \dots	& \dots	& \dots		& \dots		& 0.22/5	\\[5pt]

\bottomrule\\[-5pt]
\end{tabular}

\raggedright
\scriptsize
Joint fits of pn + MOS spectra. Errors at $1\sigma$. \\
$^{\rm a}$ Normalization of the power-law at 1 keV in units of $10^{-6}$~photons~cm$^{-2}$~s$^{-1}$~keV$^{-1}$. \\
$^{\rm b}$ With an assumed distance of 0.35 kpc.\\
$^{\rm c}$ Fixed value.\\
$^{\rm d}$ pn spectrum only.\\
\end{table*}

\section{Discussion}

Our ML analysis of the X-ray emission of four old rotation-powered pulsars has,  in general, confirmed previous results, providing in some cases smaller  uncertainties on the parameters and additional information on the spectral properties of these objects. Here, we first discuss the main results for the individual pulsars  and then we put them in the broader context of the class of old rotation-powered pulsars.

In the case of \psrc, previous analysis could not distinguish between a thermal and a non-thermal spectrum \citep{pri15}, while we found that its emission is primarily thermal, being  well fitted by a blackbody with $kT=0.17 \pm 0.02$~keV, emitting radius $R=405_{-90}^{+110}$~m, and bolometric luminosity  $L_{\rm bol} = (4.4 \pm 3.0) \x 10^{30}$ \lum. 
The small emitting area indicates that the thermal radiation comes from a hot spot of the stellar surface, most likely from the polar cap: in fact, the expected polar cap radius for \psrc is $R_{\rm PC}=((2 \pi R_*^3)/(P c))^{1/2} \approx 455$ m (for a NS radius of $R_* = 10$~km). 

We found no evidence for thermal components in the phase-averaged spectra of the other  targets and derived $3\sigma$ upper limits on their bolometric luminosity of $L_{\rm bol} \lesssim 3.2 \times 10^{28}$ \lum in \psra and $L_{\rm bol} \lesssim 2.4 \x 10^{29}$ \lum in \psrb . These limits apply for temperatures in the range $\sim\,$\band{0.05}{2} (see Fig.~\ref{fig:B06upplum} and \ref{fig:B09upplum}).  However, the  possible presence of a thermal component in  \psra is suggested by the fact that its pulsed emission is softer than the unpulsed one and it can be fitted by a blackbody  with parameters consistent, within the errors, with the above luminosity limit ($kT=0.10_{-0.01}^{+0.02}$~keV, $R=150_{-50}^{+80}$~m).

Finally, we found  features below 0.5 keV with a statistical significance of $\gtrsim\!3\sigma$ in the  spectrum of \psrd. This could be fitted  either with a power-law plus blackbody model and an absorption line at $\sim\!0.45$~keV,  or  with a power-law and  two absorption lines at harmonically spaced energies  $E_1=0.222$~keV and $E_2 = 0.444$~keV. The more realistic best fit parameters of the continuum component lead us to prefer the second interpretation.  If the lowest energy line is an electron cyclotron feature, the implied magnetic field  ($B_e=1.9 \times 10^{10} (1+z)$~G, where $z$ is the gravitational redshift), is two orders of magnitude smaller than the surface field derived from the timing parameters of the pulsar ($B_d=2.1 \times 10^{12}$~G, under the usual dipole approximation). The electrons responsible for the feature should be high in the magnetosphere, and it is unclear how they could be confined in a small region with the appropriate magnetic field value.
Alternatively, if the line is attributed to protons that could be in the atmosphere close to the NS surface, the required magnetic field is $B_p = 3.5 \times 10^{13} (1+z)$~G, implying the presence of multipolar field components, as required, for example, in the Partially Screened Gap model \citep{gil03,gil07}. 

\subsection{Thermal X-rays from hot polar caps} \label{sec:th}

\setlength{\tabcolsep}{0.5em}
\begin{table*}[htbp!]
\centering \caption{Thermal emission properties}
\label{tab:thermal}

\begin{tabular}{lcccccccc}
\toprule 
Pulsar Name & $\tau$& $R_{\rm PC}$ & $B_d$       & $d$ & $kT$ & $R_{\rm BB}$ & $L_{\rm bol}$ & Ref.\\[5pt]
            & Myr	& m            & $10^{12}$ G & kpc & keV  & m            & erg s$^{-1}$  & \\[5pt]
\midrule\\[-5pt]
B0114$+$58   & $0.28$ & $455$ & $0.8$ & $1.77\pm0.53\,^{\rm b}$ & $0.17\pm0.02$ & $405_{-90}^{+110}$ & $(4.4_{-3.8}^{+4.4})\x10^{30}$ & (1) \\[5pt]
J0633$+$1746 & $0.34$ & $297$ & $1.6$ & $0.25_{-0.08}^{+0.23}\,^{\rm a}$ & $0.16\pm0.03$ & $64\pm16$ & $(8.6_{-7.7}^{+18.})\x10^{28}$ & (2) \\[5pt]
B0656$+$14   & $0.11$ & $233$ & $4.7$ & $0.28\pm0.03\,^{\rm a}$ & $0.108\pm0.003$ & $1813\pm151$ & $(1.4\pm0.4)\x10^{31}$ & (2) \\[5pt]
B0943$+$10   & $4.98$ & $138$ & $2.0$ & $0.89\pm0.27\,^{\rm b}$ & $0.245\pm0.025$ & $35_{-8}^{+9}$ & $(1.4_{-1.2}^{+1.3})\x10^{29}$ & (3) \\[5pt]
B1055$-$52   & $0.54$ & $326$ & $1.1$ & $0.35\pm0.15\,^{\rm c}$ & $0.154\pm0.005$ & $215\pm28$ & $(8.4_{-7.5}^{+7.6})\x10^{29}$ & (2) \\[5pt]
J1740$+$1000 & $0.11$ & $369$ & $2.7$ & $1.23\pm0.37\,^{\rm b}$ & $0.148_{-0.015}^{+0.016}$ & $562_{-149}^{+237}$ & $(4.9_{-4.2}^{+6.2})\x10^{30}$ & (4) \\[5pt]
B1822$-$09   & $0.23$ & $165$ & $6.4$ & $0.3_{-0.2}^{+0.7}\,^{\rm a}$ & $0.187_{-0.023}^{+0.026}$ & $29_{-8}^{+14}$ & $(3.4_{-3.1}^{+17.0})\x10^{28}$ & (5) \\[5pt]
B1929$+$10   & $3.10$ & $304$ & $0.5$ & $0.31_{-0.06}^{+0.09}\,^{\rm a}$ & $0.30_{-0.03}^{+0.02}$ & $28_{-4}^{+5}$ & $(2.1_{-1.3}^{+1.6})\x10^{29}$ & (6) \\[5pt]
\midrule\\[-5pt]
B0355$+$54   & $0.56$ & $366$ & $0.8$ & $1.0_{-0.1}^{+0.2}\,^{\rm a}$ & $\sim\!0.16$ & $\sim\!250$ & $(1.3_{-1.1}^{+1.2})\x10^{30}$ & (7) \\[5pt]
B0628$-$28   & $2.77$ & $130$ & $3.0$ & $0.32_{-0.04}^{+0.05}\,^{\rm a}$ & $0.10_{-0.01}^{+0.02}$ & $150_{-50}^{+80}$ & $(7.3_{-5.3}^{+12.0})\x10^{28}$ & (1) \\[5pt]
B0834$+$06   & $2.97$ & $128$ & $3.0$ & $0.19\pm0.06\,^{\rm b}$ & $0.170_{-0.055}^{+0.065}$ & $7_{-3}^{+12}$ & $(1.1_{-1.0}^{+7.5})\x10^{27}$ & (8) \\[5pt]
\bottomrule\\[-5pt]
\end{tabular}

\raggedright
\scriptsize
$^{\rm a}$ Parallax measurements \citep{ver12}.\\
$^{\rm b}$ Inferred from the dispersion measure \citep{yao17}. Error assumed $30\%$.\\
$^{\rm c}$ \citet{mig10}.\\
References: (1) this paper; (2) \citet{del05}; (3) \citet{mer16}; (4) \citet{kar12a}; (5) \citet{her17}; (6) \citet{mis08}; (7) \citet{kli16}; (8) \citet{gil08}.
\end{table*}

In Table~\ref{tab:thermal}, we collected the information on the (non-recycled) rotation-powered pulsars with evidence of thermal emission from a small region of their surface. Besides the objects in which the presence of such emission is well established (first part of the table),  we have included pulsars, like \psra, for which there is only some evidence  for it (second part of the table).

The values of temperature and radius of each blackbody, with the corresponding errors, are taken from the most recent works present in literature (references in the last column of Table~\ref{tab:thermal}), and, when needed, we rescaled the radii using updated distance values.
The bolometric luminosities of these thermal components are evaluated as $L_{\rm bol}=\pi R^2 \sigma T^4$ and we included the distance uncertainties in their errors.

\begin{figure*}
	\centering
	\includegraphics[height=8cm]{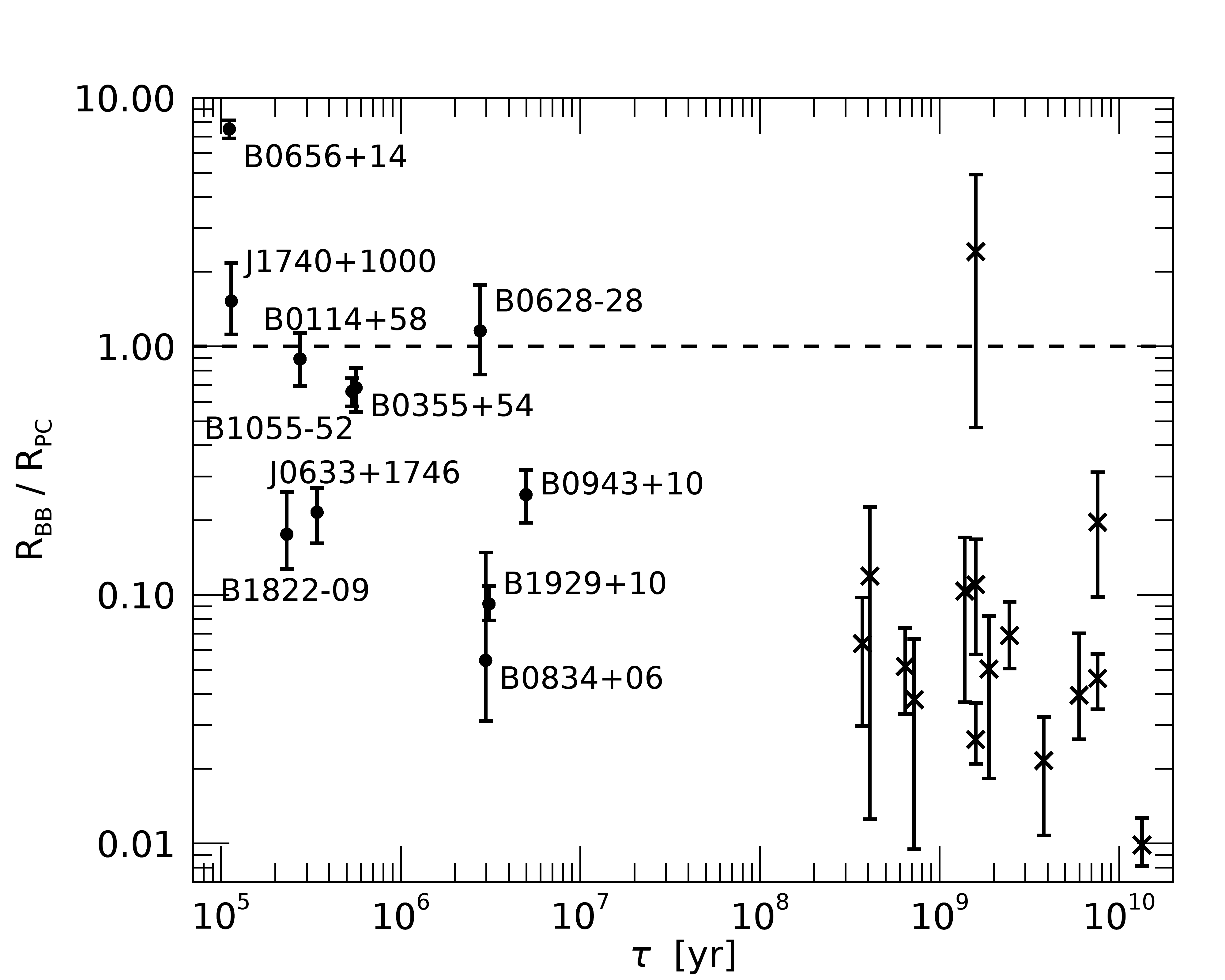}
	\caption{Ratio between the radius of emitting area inferred from the blackbody fit and the polar cap radius as a function of the pulsar characteristic age. Millisecond pulsars are indicated with a cross. In case of multiple thermal components (PSR$\,$J0030+0451 two components, PSR$\,$J0437$-$4715 three components), all the radii are plotted. \label{fig:RvsT}}
\end{figure*}

All the temperatures are in the narrow range \band{0.1}{0.3} and do not depend significantly on the  characteristic age. On the contrary, the emission radii span nearly two orders of magnitudes. We expect that the emitting region be related to the polar cap size, so we compare $R_{\rm BB}$ with the radius of the polar cap defined by the last closed lines in a dipolar field geometry, $R_{\rm PC}=((2 \pi R_*^3)/(P c))^{1/2}$.
As shown in Fig.~\ref{fig:RvsT}, in most pulsars with 0.1$<\tau<$ 10 Myr the two radii are consistent, considering the errors, and  there is no clear correlation between  $R_{\rm BB}$/$R_{\rm PC}$  and   characteristic age.
Some pulsars have $R_{\rm BB}$ significantly smaller than $R_{\rm PC}$, but this can be explained by geometrical effects.
In fact, the radii inferred from the spectral fits correspond to the projected area of the emitting region averaged over the star rotational phase.
Only for nearly aligned rotators (small angles between rotation and magnetic axis) seen pole-on  $\pi R_{\rm BB}^2$ corresponds to the real emitting area. This should occur, e.g., for PSR B0943+10 \citep{des01,bil14}, which, however, has $R_{\rm BB} < R_{\rm PC}$, possibly indicating that the magnetic field is not purely dipolar. 
In fact, the presence of multipolar components of the field can cause a reduction of the polar cap area by a factor $B_s / B_d$, where $B_s$ and $B_d$ are the actual magnetic field at the star surface and that of the dipole, respectively \citep{gil00}.
PSR$\,$B0656$+$14, on the other hand, has a blackbody radius that is nearly 10 times larger than $R_{\rm PC}$, as already noticed by \citet{del05}, suggesting the need for  a different explanation of the hot thermal component in this pulsar. One possibility is that anisotropic thermal conduction in the crust is playing a role in causing temperature gradients on the surface and the oversimplified modelling   with just two blackbodies at different temperatures give unrealistic parameters. 

\begin{figure*}
	\centering
	\includegraphics[height=8cm]{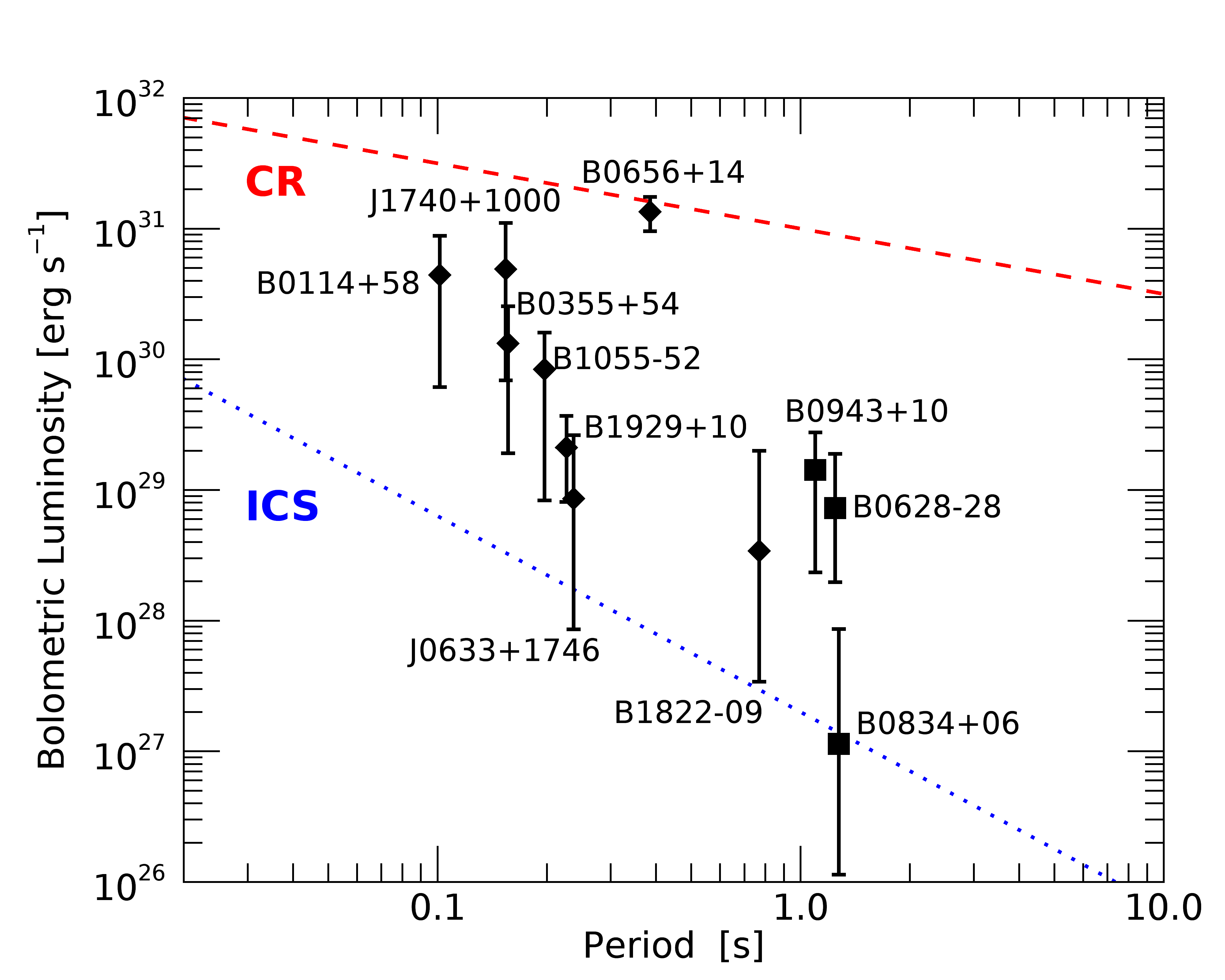}
	\caption{Bolometric luminosity of hot polar cap emission versus rotation period. The lines represent the predicted thermal luminosities in case of curvature radiation (red dashed line) and inverse-Compton scattering (blue dotted line) pair fronts. The theoretical lines are from \citet{har02a} in the case of non-recycled pulsars. Pulsars marked with a square are expected to produce pairs only from ICS photons. The error bars on the bolometric luminosity are computed taking into account both the uncertainties of the blackbody fit and those on the distances. See Table~\ref{tab:thermal} for details. \label{fig:LvsP}}
\end{figure*}

It is also interesting to compare the thermal polar cap emission of  old (non-recycled) pulsars with that observed in millisecond pulsars.
Thermal X-rays have been  detected in several millisecond pulsars \citep{bog06,zav06,spi16,bha17}. For two of them (PSR$\,$J0030+0451 \citep{bog09} and PSR$\,$J0437$-$4715 \citep{bog13}) at least two thermal components are needed, that are interpreted as emission from the polar cap having a non-uniform temperature distribution with a hot core and a cooler rim. 
We have included millisecond pulsars  in Fig.~\ref{fig:RvsT}, where, in order to make a consistent comparison with the values of non-recycled pulsars, we used the emission radii derived from blackbody fits. Note that fits with non-magnetized hydrogen atmosphere model, which are appropriate for millisecond pulsars, would result in emitting radii larger by a factor $\sim\!4$, thus making  most of them consistent with the polar caps radii in the centred dipole approximation. 

Thermal emission from small hot spots has been explored in detail by \citet{har98,har01,har02a} in the context of polar cap models.
In their ``Space Charge Limited Flow'' model, primary particles accelerated above the polar cap produce a photons-pairs cascade in an avalanche process.
Backward particles  screen the acceleration voltage drop and  heat the  polar cap.  
The pair cascade can be initiated either by curvature radiation (CR) or by inverse-Compton scattering (ICS) of thermal X-rays by primary electrons. The production  of CR photons requires a  much  higher energy of the primary particles, and therefore the pair production occurs at higher altitude and the pair heating luminosity is expected to be much higher in the CR rather than in the ICS scenario.
These processes can occur only above the corresponding ``death lines'' in the  $\pdot-P$ diagram. 
Pulsars in the region below the CR pairs death line, but above the ICS one, can produce pairs only from ICS photons.

\citet{har02a} give analytic expressions for the expected surface X-ray luminosity, depending on the pair fronts creation mechanism. In case of non-recycled pulsars, $L_{\rm bol} \approx 10^{31}$~erg s$^{-1}$ $P^{-1/2}$ for CR and $L_{\rm bol} \approx 2.0 \x 10^{27}$~erg s$^{-1}$ $P^{-3/2}$ for ICS. These relations, which should be considered as upper limits in case of incomplete screening,  are compared with the observed thermal  luminosities of the pulsars in Fig.~\ref{fig:LvsP}. Although there is a general agreement between the data and the model predictions, we note that among the three pulsars for which only the ICS mechanism should operate (indicated in Fig.~\ref{fig:LvsP} by a square), two have luminosities above the maximum predicted in this scenario.  One of them is \psra, for which the possible thermal origin of the pulsed flux is admittedly not very compelling. On the other hand, the presence of a thermal component in \psrx is well established \citep{mer16} and possibly requires a different explanation.

\section{Conclusions}
 
We applied the powerful ML technique to the spectral X-ray analysis of four dim rotation-powered old pulsars and found evidence for absorption lines at 0.22 keV and 0.44 keV in the spectrum of \psrd . If these lines are interpreted as cyclotron features due to protons they imply a magnetic field of a few 10$^{13}$ G, higher than the dipole field inferred from the timing parameters of the pulsar ($B_d = 2 \times 10^{12}$~G). 
 
The presence of thermal emission from a hot polar cap could be established only for \psrc.  Data of better statistical quality are required to ascertain if thermal components are really needed also in the other three pulsars considered here: \psra, \psrb and \psrd. Several old pulsars show the presence of thermal components in their X-ray spectra. By using the most updated spectral results and distance estimates, we found that the size of the emitting hot spots in these pulsars are generally consistent with the expected dimensions of the polar caps, once projection effects are taken into account.

\bibliographystyle{aa}
\bibliography{oldpulsar}

\begin{thebibliography}{38}
\expandafter\ifx\csname natexlab\endcsname\relax\def\natexlab#1{#1}\fi

\bibitem[{{Becker}(2009)}]{bec09}
{Becker}, W. 2009, in Astrophysics and Space Science Library, Vol. 357,
  Astrophysics and Space Science Library, ed. W.~{Becker}, 91

\bibitem[{{Becker} {et~al.}(2005){Becker}, {Jessner}, {Kramer}, {Testa}, \&
  {Howaldt}}]{bec05}
{Becker}, W., {Jessner}, A., {Kramer}, M., {Testa}, V., \& {Howaldt}, C. 2005,
  \apj, 633, 367

\bibitem[{{Bhattacharya} {et~al.}(2017){Bhattacharya}, {Heinke}, {Chugunov},
  {Freire}, {Ridolfi}, \& {Bogdanov}}]{bha17}
{Bhattacharya}, S., {Heinke}, C.~O., {Chugunov}, A.~I., {et~al.} 2017, \mnras,
  472, 3706

\bibitem[{{Bilous} {et~al.}(2014){Bilous}, {Hessels}, {Kondratiev}, {van
  Leeuwen}, {Stappers}, {Weltevrede}, {Falcke}, {Hassall}, {Pilia}, {Keane},
  {Kramer}, {Grie{\ss}meier}, \& {Serylak}}]{bil14}
{Bilous}, A.~V., {Hessels}, J.~W.~T., {Kondratiev}, V.~I., {et~al.} 2014, \aap,
  572, A52

\bibitem[{{Bogdanov}(2013)}]{bog13}
{Bogdanov}, S. 2013, \apj, 762, 96

\bibitem[{{Bogdanov} \& {Grindlay}(2009)}]{bog09}
{Bogdanov}, S. \& {Grindlay}, J.~E. 2009, \apj, 703, 1557

\bibitem[{{Bogdanov} {et~al.}(2006){Bogdanov}, {Grindlay}, {Heinke}, {Camilo},
  {Freire}, \& {Becker}}]{bog06}
{Bogdanov}, S., {Grindlay}, J.~E., {Heinke}, C.~O., {et~al.} 2006, \apj, 646,
  1104

\bibitem[{{De Luca} {et~al.}(2005){De Luca}, {Caraveo}, {Mereghetti},
  {Negroni}, \& {Bignami}}]{del05}
{De Luca}, A., {Caraveo}, P.~A., {Mereghetti}, S., {Negroni}, M., \& {Bignami},
  G.~F. 2005, \apj, 623, 1051

\bibitem[{{Deshpande} \& {Rankin}(2001)}]{des01}
{Deshpande}, A.~A. \& {Rankin}, J.~M. 2001, \mnras, 322, 438

\bibitem[{{Gil} {et~al.}(2008){Gil}, {Haberl}, {Melikidze}, {Geppert}, {Zhang},
  \& {Melikidze}}]{gil08}
{Gil}, J., {Haberl}, F., {Melikidze}, G., {et~al.} 2008, \apj, 686, 497

\bibitem[{{Gil} {et~al.}(2003){Gil}, {Melikidze}, \& {Geppert}}]{gil03}
{Gil}, J., {Melikidze}, G., \& {Geppert}, U. 2003, \aap, 407, 315

\bibitem[{{Gil} {et~al.}(2007){Gil}, {Melikidze}, \& {Zhang}}]{gil07}
{Gil}, J., {Melikidze}, G., \& {Zhang}, B. 2007, \mnras, 376, L67

\bibitem[{{Gil} \& {Sendyk}(2000)}]{gil00}
{Gil}, J.~A. \& {Sendyk}, M. 2000, \apj, 541, 351

\bibitem[{{Harding}(2013)}]{har13}
{Harding}, A.~K. 2013, Frontiers of Physics, 8, 679

\bibitem[{{Harding} \& {Muslimov}(1998)}]{har98}
{Harding}, A.~K. \& {Muslimov}, A.~G. 1998, \apj, 508, 328

\bibitem[{{Harding} \& {Muslimov}(2001)}]{har01}
---. 2001, \apj, 556, 987

\bibitem[{{Harding} \& {Muslimov}(2002)}]{har02a}
---. 2002, \apj, 568, 862

\bibitem[{{He} {et~al.}(2013){He}, {Ng}, \& {Kaspi}}]{he13}
{He}, C., {Ng}, C.-Y., \& {Kaspi}, V.~M. 2013, \apj, 768, 64

\bibitem[{{Hermsen} {et~al.}(2017){Hermsen}, {Kuiper}, {Hessels}, {Mitra},
  {Rankin}, {Stappers}, {Wright}, {Basu}, {Szary}, \& {van Leeuwen}}]{her17}
{Hermsen}, W., {Kuiper}, L., {Hessels}, J.~W.~T., {et~al.} 2017, \mnras, 466,
  1688

\bibitem[{{Hermsen} {et~al.}(2013)}]{her13}
{Hermsen}, W. {et~al.} 2013, Science, 339, 436

\bibitem[{{Kargaltsev} {et~al.}(2012){Kargaltsev}, {Durant}, {Misanovic}, \&
  {Pavlov}}]{kar12a}
{Kargaltsev}, O., {Durant}, M., {Misanovic}, Z., \& {Pavlov}, G.~G. 2012,
  Science, 337, 946

\bibitem[{{Kargaltsev} {et~al.}(2006){Kargaltsev}, {Pavlov}, \&
  {Garmire}}]{kar06}
{Kargaltsev}, O., {Pavlov}, G.~G., \& {Garmire}, G.~P. 2006, \apj, 636, 406

\bibitem[{{Klingler} {et~al.}(2016){Klingler}, {Rangelov}, {Kargaltsev},
  {Pavlov}, {Romani}, {Posselt}, {Slane}, {Temim}, {Ng}, {Bucciantini},
  {Bykov}, {Swartz}, \& {Buehler}}]{kli16}
{Klingler}, N., {Rangelov}, B., {Kargaltsev}, O., {et~al.} 2016, \apj, 833, 253

\bibitem[{{Manchester} {et~al.}(2005){Manchester}, {Hobbs}, {Teoh}, \&
  {Hobbs}}]{man05}
{Manchester}, R.~N., {Hobbs}, G.~B., {Teoh}, A., \& {Hobbs}, M. 2005, VizieR
  Online Data Catalog, 7245

\bibitem[{{Mereghetti} {et~al.}(2016){Mereghetti}, {Kuiper}, {Tiengo},
  {Hessels}, {Hermsen}, {Stovall}, {Possenti}, {Rankin}, {Esposito}, {Turolla},
  {Mitra}, {Wright}, {Stappers}, {Horneffer}, {Oslowski}, {Serylak}, \&
  {Grie{\ss}meier}}]{mer16}
{Mereghetti}, S., {Kuiper}, L., {Tiengo}, A., {et~al.} 2016, \apj, 831, 21

\bibitem[{{Mignani} {et~al.}(2010){Mignani}, {Pavlov}, \& {Kargaltsev}}]{mig10}
{Mignani}, R.~P., {Pavlov}, G.~G., \& {Kargaltsev}, O. 2010, \apj, 720, 1635

\bibitem[{{Misanovic} {et~al.}(2008){Misanovic}, {Pavlov}, \&
  {Garmire}}]{mis08}
{Misanovic}, Z., {Pavlov}, G.~G., \& {Garmire}, G.~P. 2008, \apj, 685, 1129

\bibitem[{{Prinz} \& {Becker}(2015)}]{pri15}
{Prinz}, T. \& {Becker}, W. 2015, ArXiv e-prints

\bibitem[{{Spiewak} {et~al.}(2016){Spiewak}, {Kaplan}, {Archibald}, {Gentile},
  {Hessels}, {Lorimer}, {Lynch}, {McLaughlin}, {Ransom}, {Stairs}, \&
  {Stovall}}]{spi16}
{Spiewak}, R., {Kaplan}, D.~L., {Archibald}, A., {et~al.} 2016, \apj, 822, 37

\bibitem[{{Str{\"u}der} {et~al.}(2001)}]{str01}
{Str{\"u}der}, L. {et~al.} 2001, \aap, 365, L18

\bibitem[{{Szary} {et~al.}(2017){Szary}, {Gil}, {Zhang}, {Haberl}, {Melikidze},
  {Geppert}, {Mitra}, \& {Xu}}]{sza17}
{Szary}, A., {Gil}, J., {Zhang}, B., {et~al.} 2017, \apj, 835, 178

\bibitem[{{Tepedelenl{\i}o{\v g}lu} \& {{\"O}gelman}(2005)}]{tep05}
{Tepedelenl{\i}o{\v g}lu}, E. \& {{\"O}gelman}, H. 2005, \apjl, 630, L57

\bibitem[{{Turner} {et~al.}(2001)}]{tur01}
{Turner}, M.~J.~L. {et~al.} 2001, \aap, 365, L27

\bibitem[{{Verbiest} {et~al.}(2012){Verbiest}, {Weisberg}, {Chael}, {Lee}, \&
  {Lorimer}}]{ver12}
{Verbiest}, J.~P.~W., {Weisberg}, J.~M., {Chael}, A.~A., {Lee}, K.~J., \&
  {Lorimer}, D.~R. 2012, \apj, 755, 39

\bibitem[{{Yao} {et~al.}(2017){Yao}, {Manchester}, \& {Wang}}]{yao17}
{Yao}, J.~M., {Manchester}, R.~N., \& {Wang}, N. 2017, \apj, 835, 29

\bibitem[{{Zavlin}(2006)}]{zav06}
{Zavlin}, V.~E. 2006, \apj, 638, 951

\bibitem[{{Zharikov} \& {Mignani}(2013)}]{zha13}
{Zharikov}, S. \& {Mignani}, R.~P. 2013, \mnras, 435, 2227

\bibitem[{{Zharikov} {et~al.}(2008){Zharikov}, {Shibanov}, {Mennickent}, \&
  {Komarova}}]{zha08}
{Zharikov}, S.~V., {Shibanov}, Y.~A., {Mennickent}, R.~E., \& {Komarova}, V.~N.
  2008, \aap, 479, 793

\end{thebibliography}
   
\end{document}